\documentclass[unnumsec,webpdf,contemporary,large]{oup-authoring-template}%

\usepackage{float}
\usepackage{subcaption}
\usepackage{booktabs} 
\usepackage{graphicx}
\usepackage{hyperref}
\usepackage{array}
\usepackage{colortbl}
\usepackage{xcolor}
\usepackage{caption}
\usepackage{float}

\theoremstyle{thmstyleone}%
\theoremstyle{thmstyletwo}%
\theoremstyle{thmstylethree}%

\begin{document}

\journaltitle{}
\DOI{DOI HERE}
\copyrightyear{2024}
\pubyear{2024}
\access{Advance Access Publication Date: Day Month Year}
\appnotes{Reflection}

\firstpage{1}


\title[Topological convergence in Bayesian phylogenetics]{On the importance of assessing topological convergence in Bayesian phylogenetic inference}

\author[1,$\ast$]{Marius Brusselmans}
\author[2]{Luiz Max Carvalho}
\author[1]{Samuel L. Hong}
\author[3]{Jiansi Gao}
\author[4,5,6]{Frederick A. Matsen IV\ORCID{0000-0003-0607-6025}}
\author[7]{Andrew Rambaut}
\author[1]{Philippe Lemey}
\author[8]{Marc A. Suchard}
\author[9]{Gytis Dudas}
\author[1]{Guy Baele\ORCID{0000-0002-1915-7732}}

\authormark{Brusselmans et al.}
\address[1]{\orgdiv{Department of Microbiology, Immunology and Transplantation}, \orgname{Rega Institute}, 
\orgaddress{
\postcode{KU Leuven}, \state{Leuven}, \country{Belgium}}}

\address[2]{\orgdiv{School of Applied Mathematics, \orgname{Getulio Vargas Foundation}, \orgaddress{\street{Praia de Botafogo, 190}, \postcode{22250-900}}, \state{Rio de Janeiro}, \country{Brazil}}}

\address[3]{\orgdiv{Computational Biology Program}, \orgname{Fred Hutchinson Cancer Center},
 \orgaddress{
\postcode{Seattle}, \state{WA 98109}, \country{USA}}}

\address[4]{\orgdiv{Howard Hughes Medical Institute, Computational Biology Program}, \orgname{Fred Hutchinson Cancer Research Center},
 \orgaddress{
\postcode{Seattle}, \state{Washington}, \country{USA}}}

\address[5]{\orgdiv{Department of Genome Sciences}, \orgname{University of Washington},
 \orgaddress{
\postcode{Seattle}, \state{Washington}, \country{USA}}}

\address[6]{\orgdiv{Department of Statistics}, \orgname{University of Washington},
 \orgaddress{
\postcode{Seattle}, \state{Washington}, \country{USA}}}

\address[7]{\orgdiv{Institute of Ecology and Evolution}, \orgname{University of Edinburgh},
 \orgaddress{
\postcode{Edinburgh}, \state{EH9, 3FL}, \country{UK}}}

\address[8]{\orgdiv{Department of Biostatistics, Fielding School of Public Health}, \orgname{University of California},
 \orgaddress{
\postcode{Los Angeles}, \state{CA 90095}, \country{USA}}}

\address[9]{\orgdiv{Institute of Biotechnology}, \orgname{Life Sciences Center},
 \orgaddress{
\postcode{Vilnius University}, \state{Vilnius}, \country{Lithuania}}}

\corresp[$\ast$]{Corresponding author. \href{email: marius.brusselmans@kuleuven.be}{marius.brusselmans@kuleuven.be}}

\received{Date}{0}{Year}
\revised{Date}{0}{Year}
\accepted{Date}{0}{Year}


\abstract{Modern phylogenetics research is often performed within a Bayesian framework, using sampling algorithms such as Markov chain Monte Carlo (MCMC) to approximate the posterior distribution.
These algorithms require careful evaluation of the quality of the generated samples.
Within the field of phylogenetics, one frequently adopted diagnostic approach is to evaluate the \textit{effective sample size} (ESS) and to investigate trace graphs of the sampled parameters.
A major limitation of these approaches is that they are developed for continuous parameters and therefore incompatible with a crucial parameter in these inferences: the \textit{tree topology}.
Several recent advancements have aimed at extending these diagnostics to topological space.
In this reflection paper, we present two case studies -- one on Ebola virus and one on HIV -- illustrating how these topological diagnostics can contain information not found in standard diagnostics, and how decisions regarding which of these diagnostics to compute can impact inferences regarding MCMC convergence and mixing.
Our results show the importance of running multiple replicate analyses and of carefully assessing topological convergence using the output of these replicate analyses.
To this end, we illustrate different ways of assessing and visualizing the topological convergence of these replicates.
Given the major importance of detecting convergence and mixing issues in Bayesian phylogenetic analyses, the lack of a unified approach to this problem warrants further action, especially now that additional tools are becoming available to researchers.
}

\keywords{effective sample size, topologies, Bayesian inference, phylogenetics, phylodynamics, convergence, mixing, EBOV, HIV}

\maketitle

\section{Introduction}
\subsection*{Background}
When performing Bayesian phylogenetic inference using Markov chain Monte Carlo (MCMC) algorithms, the standard practice to evaluate the quality of the generated samples is to visually inspect trace plots and to compute the corresponding effective sample size (ESS) of the sampled parameters.
This can be done using a variety of software packages such as \textit{Tracer} \citep{tracer}, \textit{Beastiary} \citep{beastiary} or \textit{CODA} \citep{coda}, for example.
However, such software packages only produce diagnostics for simple, often univariate and continuous, model parameters, which the topology of the phylogenetic tree is not. 
If these diagnostics suggest the MCMC convergence and mixing of the simple model parameters are satisfactory, it is usually assumed that this will be the case for the topology as well.
This is potentially problematic, as the tree topology is often of key interest in phylogenetic and phylodynamic studies, and obtaining a correct (consensus) phylogeny is essential when performing outbreak investigation and monitoring ongoing epidemics (see e.g. \citet{attwood}).

Recent research has focused on convergence and mixing diagnostics applicable to the tree topology, including studies by \citet{lanfear}, \citet{Magee}, and \citet{Hohna}.
The former two studies focus on finding ways to apply the principles of trace graphs and effective sample sizes to the topology as a whole, while the latter study considers the presence of each split in the tree as an individual parameter to be evaluated using classical diagnostics.
We here explore the main topological methods introduced by \citet{lanfear} and \citet{Hohna}, as well as those from \citet{Magee} on the data from an Ebola virus (EBOV) study by \citet{ebola} and an HIV study by \citet{hiv}.
Of note, we also include the multidimensional scaling (MDS) ESS metric -- a more conservative tree ESS measure \citep{Magee} -- that also enables us to project the high-dimensional phylogenies onto a small number of dimensions suitable for visualisation \citep{Kruskal64a,Kruskal64b}.
 
We find that the evaluation of convergence and mixing of samples in topological space can reveal issues not typically detected by standard diagnostics for (convergence and mixing of) continuous parameters.
Furthermore, we find that decisions regarding the computation of these diagnostics can impact the conclusions regarding MCMC convergence and mixing.
We selected the EBOV study because of its large size and the rich complexity of the models that were applied, making it a prime case study of a challenging phylogenetic analysis that could be susceptible to hidden convergence and mixing issues.
Further, HIV phylogenies are known to be star-like (i.e., have short internal but long external branches), which could lead to a different set of issues from a topological perspective.

\subsection*{Phylogenetic distance metrics}

A central concept in our exploration is that of phylogenetic distance, a quantitative measure of similarity between two phylogenetic trees.
Such a distance can be computed in several ways, using what we will refer to as phylogenetic distance metrics.
These distances are zero for two identical trees and are expected to increase as trees grow more dissimilar.
The metrics considered in this paper are: the \textbf{Robinson-Foulds} distance \citep{RFdist} and its weighted counterpart \citep{wRF}, the \textbf{path difference} \citep{PD}, the \textbf{branch score} \citep{KF}, the \textbf{Kendall-Colijn} distance \citep{KC} with $\lambda$=0 (i.e., disregarding the branch lengths), and the rooted \textbf{subtree-prune-regraft (SPR) distance} \citep{RSPR}. 

In order to provide the reader with some intuition of what aspects of topological differences these metrics convey, we use a toy example of two phylogenetic trees with the same set of time-calibrated taxa on which an example calculation of each of these metrics is performed.

\begin{figure}[!htbp]
    \centering
    \begin{minipage}{\linewidth}
        \centering
        \includegraphics[width=0.7\linewidth]{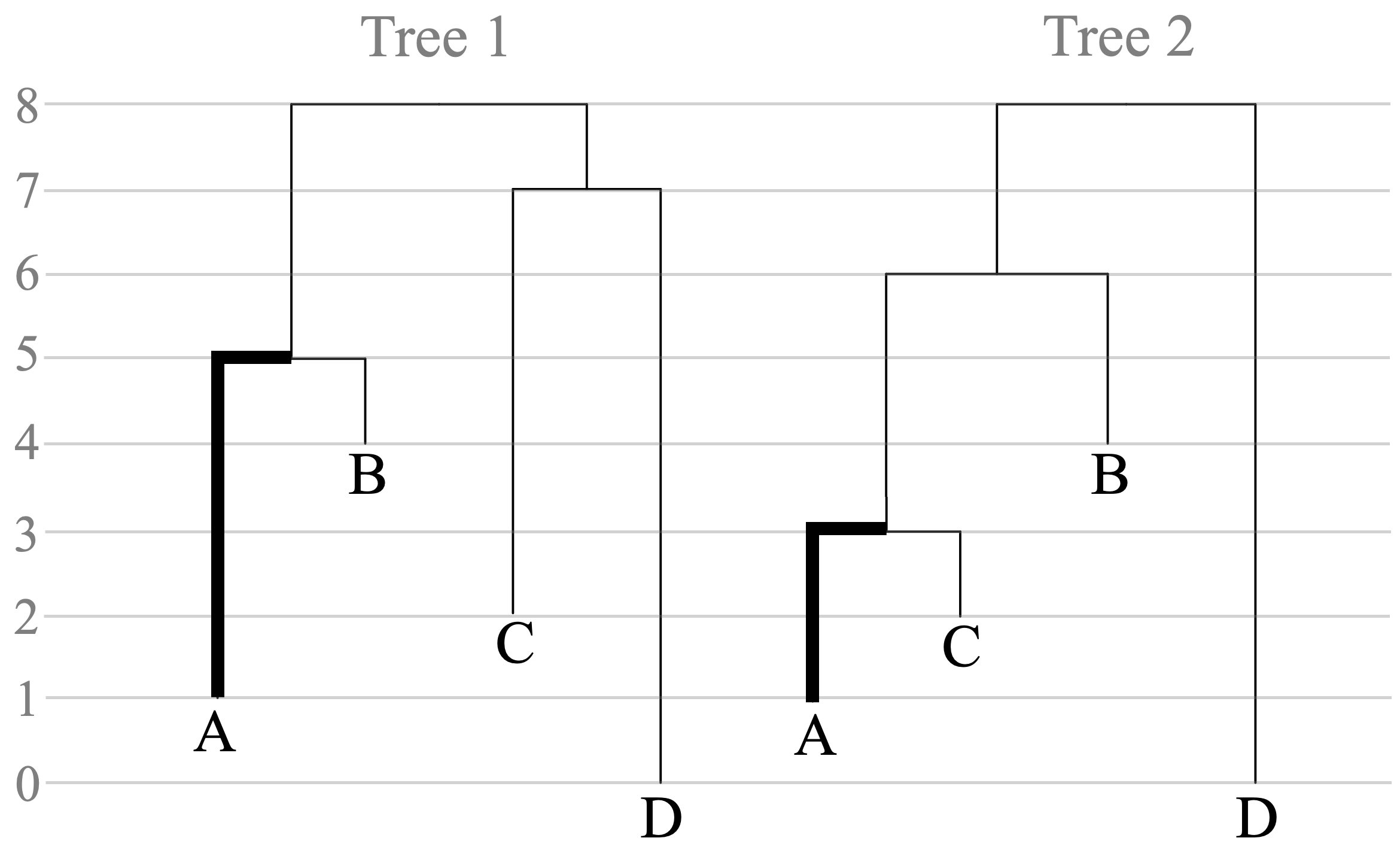} 
		\vspace{.3cm}
    \end{minipage}
    \begin{minipage}{\linewidth}
        \centering
        \begin{tabular}{*6c}
			\toprule
			Partition &  \multicolumn{3}{c}{Associated branch lengths} \\ \cline{2-4}
			 &  tree 1 &  tree 2 & $\vert\Delta\vert$ \\
			\midrule
			\textbf{\{A$\vert$BCD\}} & \textbf{4} & \textbf{2} & \textbf{2}\\
			\{B$\vert$ACD\} & 1 & 2 & 1\\
			\{C$\vert$ABD\} & 5 & 1 & 4\\
			\{D$\vert$ABC\} & 7 & $8+2$ & 3\\
			\{AB$\vert$CD\} & $3+1$ & \textit{none} & 4\\
			\{AC$\vert$BD\} & \textit{none} & 3 & 3\\
			\bottomrule
		\end{tabular}
		\vspace{.3cm}
        \caption{Partitions defined by each branch in the two trees and the associated branch lengths. The branches defining partition \{A$\vert$BCD\} and their associated branch lengths are \textbf{bolded} in both trees and the table. $\vert\Delta\vert$ is the absolute difference of two corresponding branch lengths. Certain partitions do not exist in one of either trees, in which case the branch length is shown as \textit{none} and treated as 0 for the computation of $\vert\Delta\vert$. The \textbf{(weighted) Robinson-Foulds distance} and the \textbf{branch score} can be computed from the information contained in the provided table.}
		\label{figure:splitdiffs}
    \end{minipage}
\end{figure}

Figure~\ref{figure:splitdiffs} shows, for each branch in the tree, the partition defined by that branch.
Each of these branches also has an associated branch length in both trees, which is reported as well.
The Robinson-Foulds distance is simply the number of partitions present in one tree but not the other, which in this case is 2, since only \{AB$\vert$CD\} and \{AC$\vert$BD\} are such partitions.
The weighted Robinson-Foulds distance would be the sum of the absolute differences in the branch lengths of branches defining corresponding partitions, which are shown in the $\vert\Delta\vert$ column. This distance is thus 17. Closely related is the branch score, which takes the square root of the sum of squares of $\vert\Delta\vert$, which in this case equals $7.42$.

\begin{figure}[!htbp]
    \centering
    \begin{minipage}{\linewidth}
        \centering
        \includegraphics[width=0.7\linewidth]{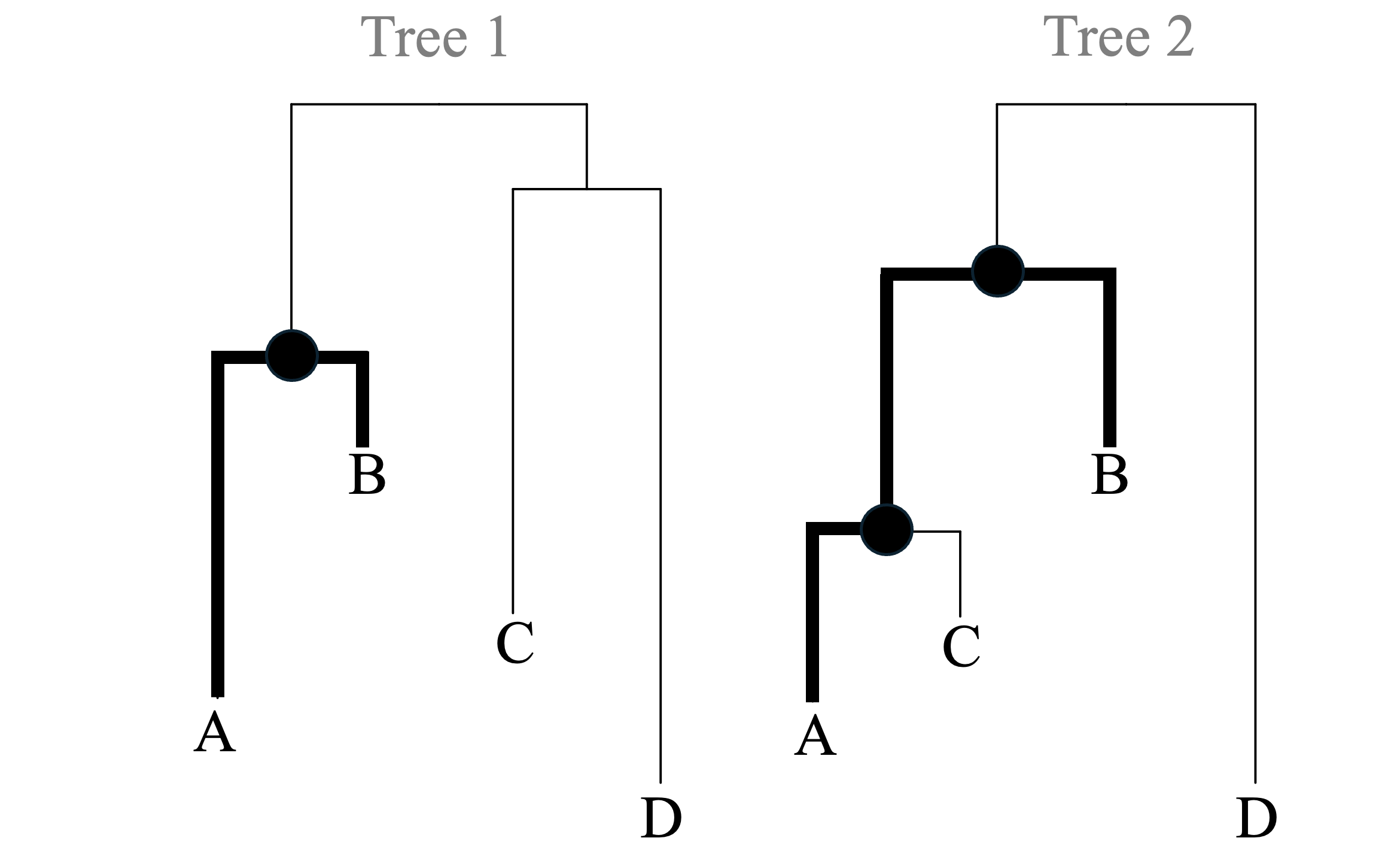} 
		\vspace{.3cm}
    \end{minipage}
    \begin{minipage}{\linewidth}
        \centering
        \begin{tabular}{*7c}
			\toprule
			Pair of tips &  \multicolumn{3}{c}{$d_{P}$} &  \multicolumn{3}{c}{$d_{MRCA-root}$}\\
			 \cline{2-4} \cline{5-7}
			 &  tree 1 &  tree 2 & $\vert\Delta\vert$ &  tree 1 &  tree 2 &  $\vert\Delta\vert$ \\			
			 \midrule
				\textbf{\{A-B\}} & \textbf{1} & \textbf{2} & 1 & 1 & 1 & 0\\
				\{A-C\} & 2 & 1 & 1 & 0 & 2 & 2\\
				\{A-D\} & 2 & 2 & 0 & 0 & 0 & 0\\
				\{B-C\} & 2 & 2 & 0 & 0 & 1 & 1\\
				\{B-D\} & 2 & 1 & 1 & 0 & 0 & 0\\
				\{C-D\} & 1 & 2 & 1 & 1 & 0 & 1\\
			\bottomrule
		\end{tabular}
		\vspace{.3cm}
        \caption{Pairwise tip-to-tip path lengths $d_{P}$, defined as the number of internal nodes that must be crossed (not counting the root node) to get from one tip to another, as well as absolute difference between these. The path to go from tip $A$ to $B$ is \textbf{bolded} in both trees, as are the crossed internal nodes. Also shown is the closely related MRCA-to-root path length $d_{MRCA-root}$ for each pair of tips, defined as the number of internal nodes that must be crossed to go from the MRCA of the tips to the root node (MRCA node included). The \textbf{path difference} and \textbf{Kendall-Colijn} distance can be computed from the information contained in the provided table.}
	\label{figure:pathdiffs}
    \end{minipage}
\end{figure}

Figure~\ref{figure:pathdiffs} shows - for both trees - all pairwise tip-to-tip path lengths, which is the number of internal nodes that must be crossed to go from one tip to the other.
The path difference is the square root of the sum of squares of the differences in path lengths $d_{P}$ between the two trees, which in this case equals 2.
Figure~\ref{figure:pathdiffs} also shows the path lengths from the most recent common ancestor (MRCA) of each pair of tips to the root node $\vert\Delta\vert$.
The Kendall-Colijn distance (with $\lambda=0$, as considered in this manuscript) is the square root of the sum of squares of these $\vert\Delta\vert$ values, which in this case equals 2.45.

Thus, these distance metrics can be divided into two general categories. First are metrics defined by partitions and branch lengths (Robinson-Foulds, weighted Robinson-Foulds, branch score) as shown in Figure~\ref{figure:splitdiffs}. Second are metrics defined by path lengths between tips and/or nodes (path difference, Kendall-Colijn).

\begin{figure}[!htbp]
	\centering
	\includegraphics[width=\linewidth]{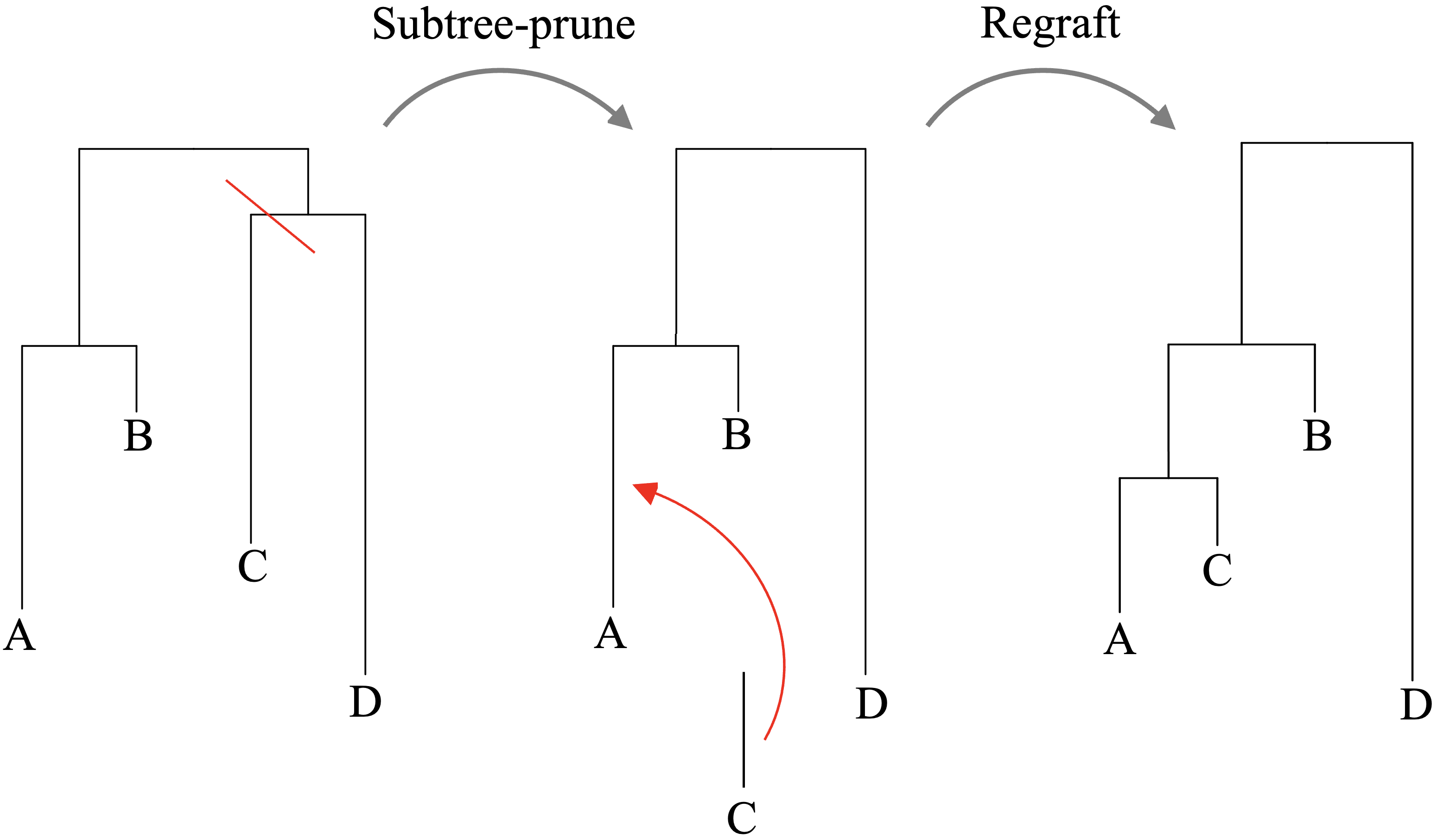}
	\caption{The subtree-prune-regraft (SPR) move required to transform tree 1 into tree 2, disregarding branch lengths. The move involves pruning the branch defining the \{C$\vert$ABD\} partition and regrafting it onto the branch defining the \{A$\vert$BD\} partition. The SPR distance is the number of SPR moves required to transform tree 1 into tree 2, in this case 1.}
	\label{figure:sprmove}
\end{figure}

A last type of distance metric we consider is the subtree-prune-regraft (SPR) distance. 
SPR refers to a type of operation that can be performed on a tree to change its topology.
This is closely related to the nature of the trees considered in this manuscript, since most Bayesian phylogenetic software packages make use of SPR-like moves to explore tree space.
Figure~\ref{figure:sprmove} shows the SPR move that would be required to transform tree 1 into tree 2.
Since only one such move is needed, the SPR distance between the two trees shown equals 1.

\subsection*{Topological convergence diagnostics} 

We here briefly discuss the topological convergence diagnostics considered in this manuscript.

The \textbf{topology trace plot}, as defined by \citet{lanfear}, works very similarly to standard trace plots used for continuous parameters.
The value of the trace being graphed is the phylogenetic distance from each sampled tree to a chosen reference tree.
As a reference tree, one can choose between, for example, one of the posterior trees or a consensus tree. 
However, choosing a tree that is part of the chain will cause a ``slump'' in the trace plot as the distance from this tree to itself is inevitably equal to zero. 
It is good practice to try several reference trees and compare the resulting trace plots. 
In this manuscript, we use the first tree as a reference and exclude it from the graphs to avoid scaling issues. 

The \textbf{pseudo-ESS} \citep{lanfear} is the ESS of the vector of phylogenetic distances from an arbitrarily chosen focal tree to all other trees in the sample (note the analogy with the topology trace plot).
Because there is randomness involved in the choice of the focal tree, the computation is repeated using each tree in the sample as a focal tree.
The lowest value and median value are then reported.

The \textbf{approximate ESS} \citep{lanfear} derives an approximation of the common ESS calculation of dividing the actual sample size by the autocorrelation time, but uses a topological version of autocorrelation time estimated by determining the thinning interval at which the average phylogenetic distance between subsequently sampled trees ceases to increase as the thinning interval increases. 

The \textbf{Fréchet correlation ESS} \citep{Magee} is defined analogously to the standard one-dimensional continuous ESS, but substitutes Pearson autocorrelation for an alternative definition of autocorrelation between trees using Fréchet (co)variances, which make use of the relationship between covariance and Euclidian distance -- here substituted with whatever phylogenetic distance is being used. 

The \textbf{split frequency ESS} \citep{Magee} is computed by treating each tree as a vector of binary split indicators (split is present/absent).
Fréchet (co)variances can then be computed using the Euclidean norm (as the trees are now reduced to vectors of binary indicators), which enable computation of an ESS.
Note that this is the only method that does not explicitly use any kind of phylogenetic distance metric.

The \textbf{multidimensional scaling ESS} \citep{Magee} (MDS ESS) is computed by performing multidimensional scaling of the matrix of squared pairwise distances between trees.
The first dimension of the resulting multidimensional scaling matrix is then used to compute an ESS on.

\section{Materials and Methods}

\subsection*{Software}

We computed the topological ESS estimators described in the previous section using the \texttt{treess} package version 1.0.1 \citep{Magee} with R v4.3.0 \citep{R}, the phylogenetic distances required for these ESS estimators using the R packages \texttt{phangorn} v2.11.1 \citep{phangorn} and \texttt{TreeDist} v2.6.1 \citep{treedist}, the per-split ESS values with the R package \texttt{convenience} \citep{Hohna}, and the approximate subtree-prune-regraft (aSPR) distances using RSPR version 1.3.1 \citep{RSPR}.
We constructed the tanglegrams using Baltic version 0.2.2, available at \url{https://github.com/evogytis/baltic}.
Finally, we also made use of \textit{Tracer} v1.7.2 \citep{tracer} and the \textit{TreeAnnotator} v1.10.4 tool associated with the BEAST 1.10.4 software package for summarising maximum clade credibility (MCC) trees \citep{beast}, as well as the R package \texttt{ggtree} v3.8.2 \citep{ggtree} for visualisation of these trees. 

\subsection*{Data}

The first dataset considered in this study is from a genomic epidemiology study of the largest Ebolavirus (EBOV) outbreak to date, which investigated the impact of several potential predictors, such as climate and demographic information, on EBOV spread in West Africa from 2014 to 2015.
We refer the interested reader to the original publication for further details \citep{ebola}. 
In summary, \citet{ebola} performed Bayesian phylogenetic inference using MCMC on a total of 1\,610 EBOV genome sequences sampled between March 2014 and October 2015 using an HKY + $\Gamma_4$ nucleotide substitution model, an uncorrelated relaxed molecular clock model with an underlying lognormal distribution and a flexible non-parametric coalescent model \citep{skygrid} as the tree prior.
We downloaded the 1\,000 posterior sample trees, which were sampled every 10\,000 iterations, and log-files from the original publication from \url{https://github.com/ebov/space-time/tree/master/Analyses/Phylogenetic}.
Of note, the burn-in was already discarded from the posterior sample trees file shared by \citet{ebola}.

We obtained the second dataset considered from a 2020 phylogeographic study of the spread of HIV-1 subtype B in the USA.
As in the EBOV study, the HIV study also aimed to identify relevant covariates for the spread of the virus, but also studied the impact of different subsampling schemes on these inferences.
We refer interested readers to the original publication for further details \citep{hiv}, from which we selected a data set consisting of 500 sequences that was constructed with the aim of maximizing phylogenetic diversity.
The authors used a GTR + $\Gamma_4$ nucleotide substitution model, a strict molecular clock model, and a logistic population growth model acting as a tree prior.
The output of the original analysis also consists of 1\,000 posterior sample trees.
The burn-in was already discarded from the posterior sample trees file, which is available on \url{https://github.com/hongsamL/HIV_trees}.

\section{Assessing topological convergence}

\subsection*{The EBOV data}

\begin{figure*}[!h]
	\begin{center}
	\includegraphics[scale=0.28]{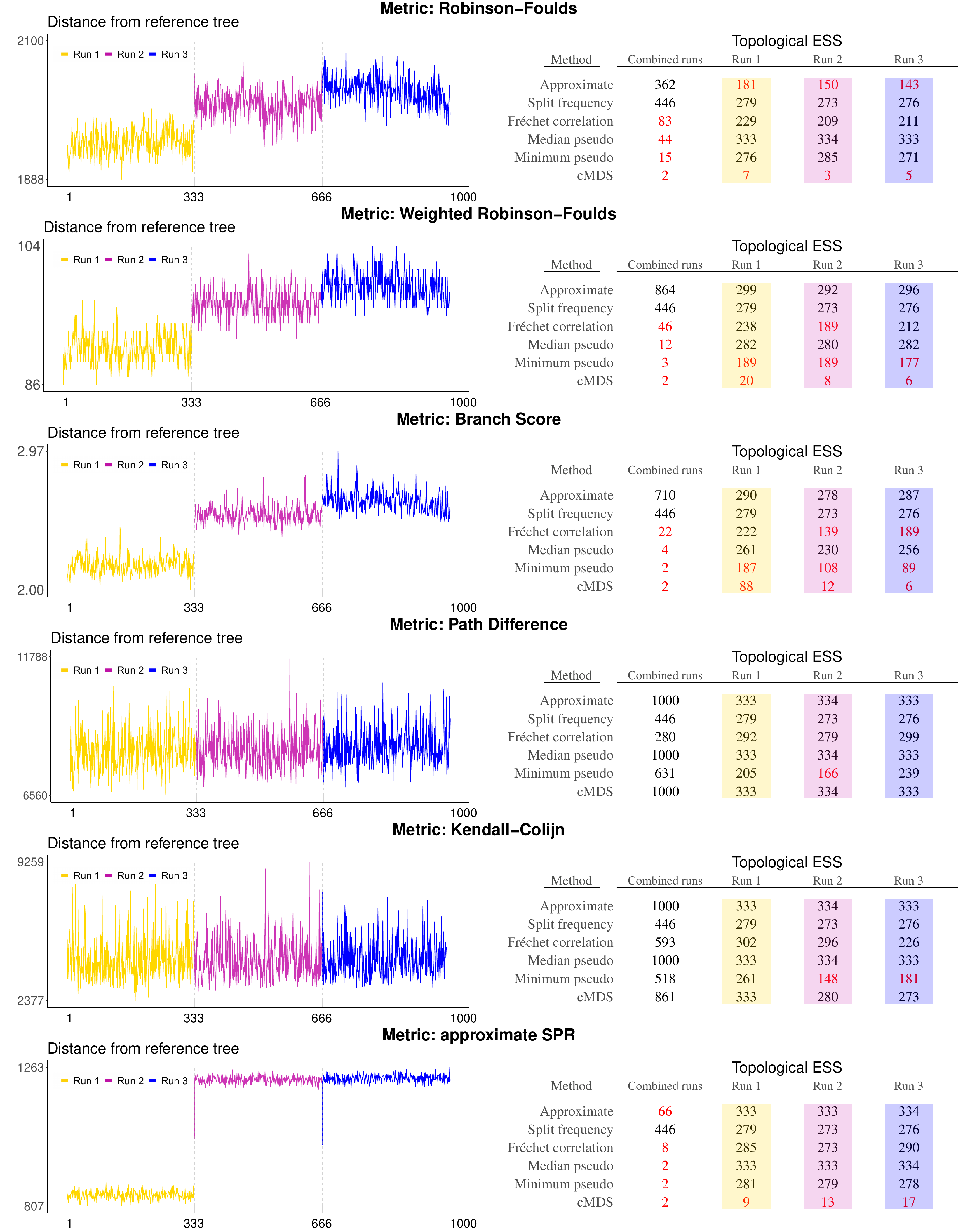}
	\caption{Topology trace plots for the EBOV analysis using six different phylogenetic distance metrics, with associated topological ESS estimates. Color-coded trace plots correspond to color-coded columns in the topological estimates, with all of them generated from samples from the same three independent BEAST runs. The X-axis shows the index of the sampled tree (thinned sample), while the Y-axis shows the phylogenetic distance from the first tree. The distance from the first tree to itself (which is inevitably 0) is excluded from the graph to avoid scaling issues. ESS values below 200---an often-used cutoff in practice for terminating running analyses in Bayesian phylogenetics---are highlighted in red. This sample of 1\,000 trees is in fact a concatenated set of three samples from independent \textit{BEAST} runs that yielded 333, 333, and 334 trees respectively. Notice how certain topological distance metrics show clear jumps between the separate runs, while for others the traces seem entirely homogeneous.
	}
	\label{fig:toptraceebola}
	\end{center}
\end{figure*}

Figure~\ref{fig:toptraceebola} shows the topology trace plots for the EBOV analysis using the six different topological distance metrics, as well as the corresponding topological ESS estimates. 
The results in this figure can be grouped in two sets based on the distance metric used: the (weighted) Robinson-Foulds distance, the branch score and the aSPR distance on the one hand, and the path difference and Kendall-Colijn distance on the other.
Note that this grouping of metrics is closely related to the conceptual differences between them as shown in Figures~\ref{figure:splitdiffs} and \ref{figure:pathdiffs}.

In the former group of metrics, the combined traces can be clearly divided into three distinct parts at iterations 333 and 666. 
This suggests that three independent replicate analyses were combined to obtain the posterior sample of trees in the study of \citet{ebola}, which has been confirmed by the authors.
Bayesian phylogenetic inference on large data sets indeed commonly employs the practice of concatenating the samples of several independent chains to both reduce computation time (by increasing the ESS values of continuous parameters) and assess convergence towards the same posterior.
Four of the topology trace plots in Figure~\ref{fig:toptraceebola} therefore suggest a discrepancy between the posterior space explored by the three chains.
Whether this indicates failure to converge to the same posterior, a case of extremely slow / poor mixing, or something else entirely is not clear.
The ESS estimates in this group tend to be substantially lower when computed for the entire sample than when computed for the three individual samples, which can be expected if the subsamples explore differents parts of tree space.
This is not entirely consistent though, as the approximate ESS only shows this behaviour when considering the aSPR distance, and the split-frequency ESS does not show this behaviour at all.
It should be remarked that the split frequency ESS is invariant to the choice of phylogenetic distance metric, as it is computed on the vector of splits directly.

In the latter group of metrics---considering the path difference and the Kendall-Colijn distance---the three individual samples are indistinguishable in the topology trace plots.
The ESS values are also substantially better accross the board and do not decrease when considering the whole concatenated sample as opposed to the individual samples, but instead are higher than the ESS of each individual sample.

\subsection*{The HIV data}

\begin{figure*}[!h]
	\begin{center}
	\includegraphics[scale=0.28]{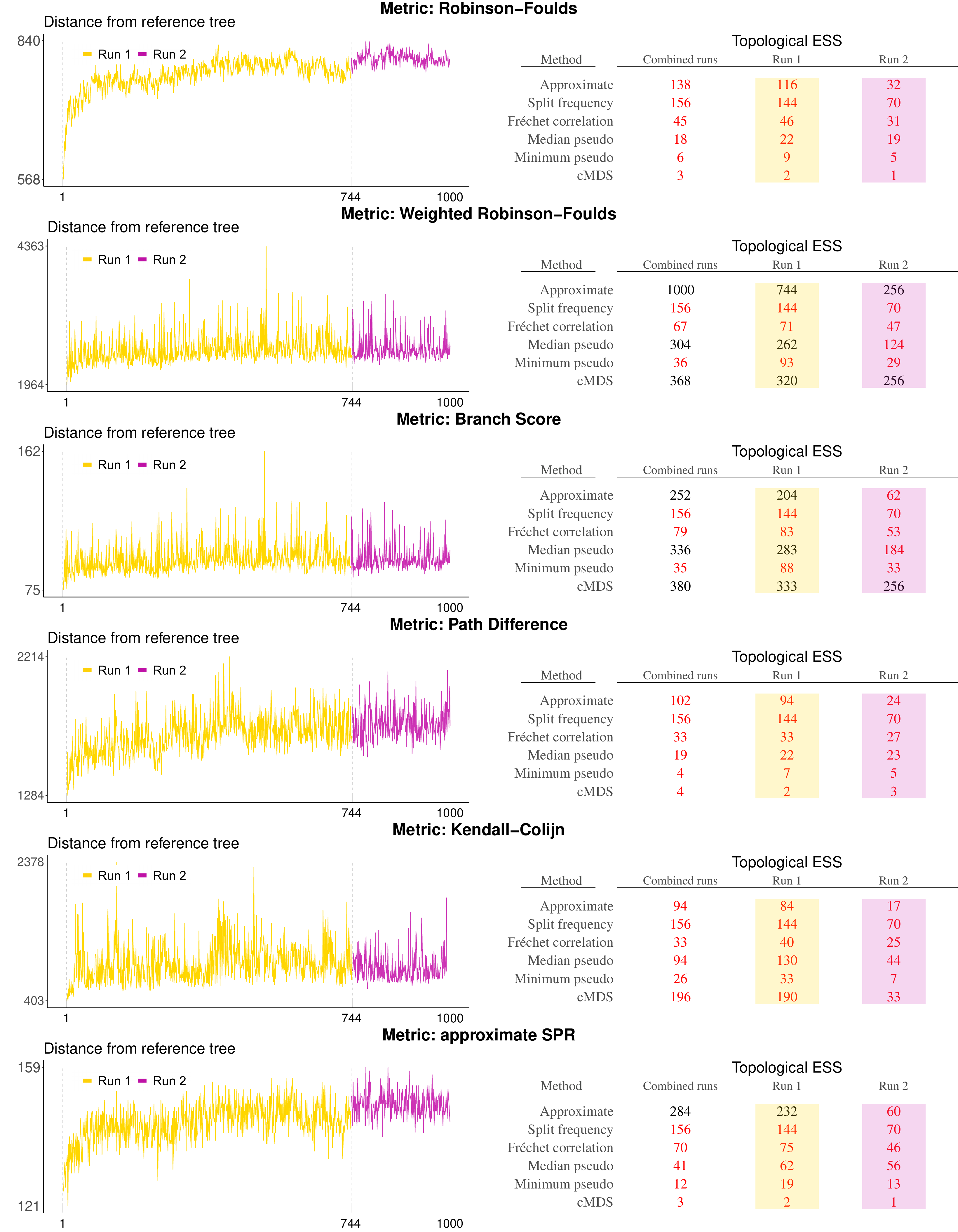}
	\caption{Topology trace plots for the HIV analysis using six different phylogenetic distance metrics, with associated topological ESS estimates. Color-coded trace plots correspond to color-coded columns in the topological estimates, with all of them generated from samples from the same two independent BEAST runs. The X-axis shows the index of the sampled tree (thinned sample), while the Y-axis shows the phylogenetic distance from the first tree. The distance from the first tree to itself (which is inevitably 0) is excluded from the graph to avoid scaling issues. ESS values below 200---an often used cutoff in practice for terminating running analyses in Bayesian phylogenetics---are coloured in red. This sample of 1\,000 trees is in fact a concatenated set of two samples from distinct \textit{BEAST} runs that yielded 744 and 256 trees respectively.
	}
	\label{fig:toptracehiv}
	\end{center}
\end{figure*}

Figure~\ref{fig:toptracehiv} shows the topology trace plots for the HIV analysis using the six different topological distance metrics, as well as the corresponding topological ESS estimators.
Similar to the EBOV analysis, the full sample is a concatenated set of two samples (as confirmed by the authors of \citep{hiv}), but with each sample containing a different number of trees.
However, unlike with the EBOV data, this is not immediately apparent from any of the topology trace plots (of the combined sample). 
All distance metrics show a clear slump at the beginning, which could be indicative of the discarded burn-in not having been set sufficiently high from a topological perspective (but indeed set sufficiently high from the current standard practice of only assessing the traces of continuous parameters), which suggests part of the sample is from a non-converged part of the analysis. 
Topological ESS values tend to be substantially lower across the board for this dataset, most often not even reaching the often-used ESS cut-off of 200 for continuous parameters.
Further, the difference in behaviour between the different distance metrics as seen in the EBOV data is not apparent here.

\section{Alternative visualisations}

The topological trace graphs presented in the previous section require relatively few computational resources to generate, but only provide partial information as they only consider distances from a single reference tree.
We present two alternative kinds of visualisations - generated from pairwise distance calculations (meaning all distances between all possible pairs of the 1\,000 trees in the samples: producing 449\,500 distances) to produce a more comprehensive visual mapping of topological distances between sampled trees.

\subsection*{Pairwise heatmaps}

\begin{figure}[!h]
	\begin{center}
	\begin{subfigure}[b]{2in}
		\includegraphics[scale=0.08]{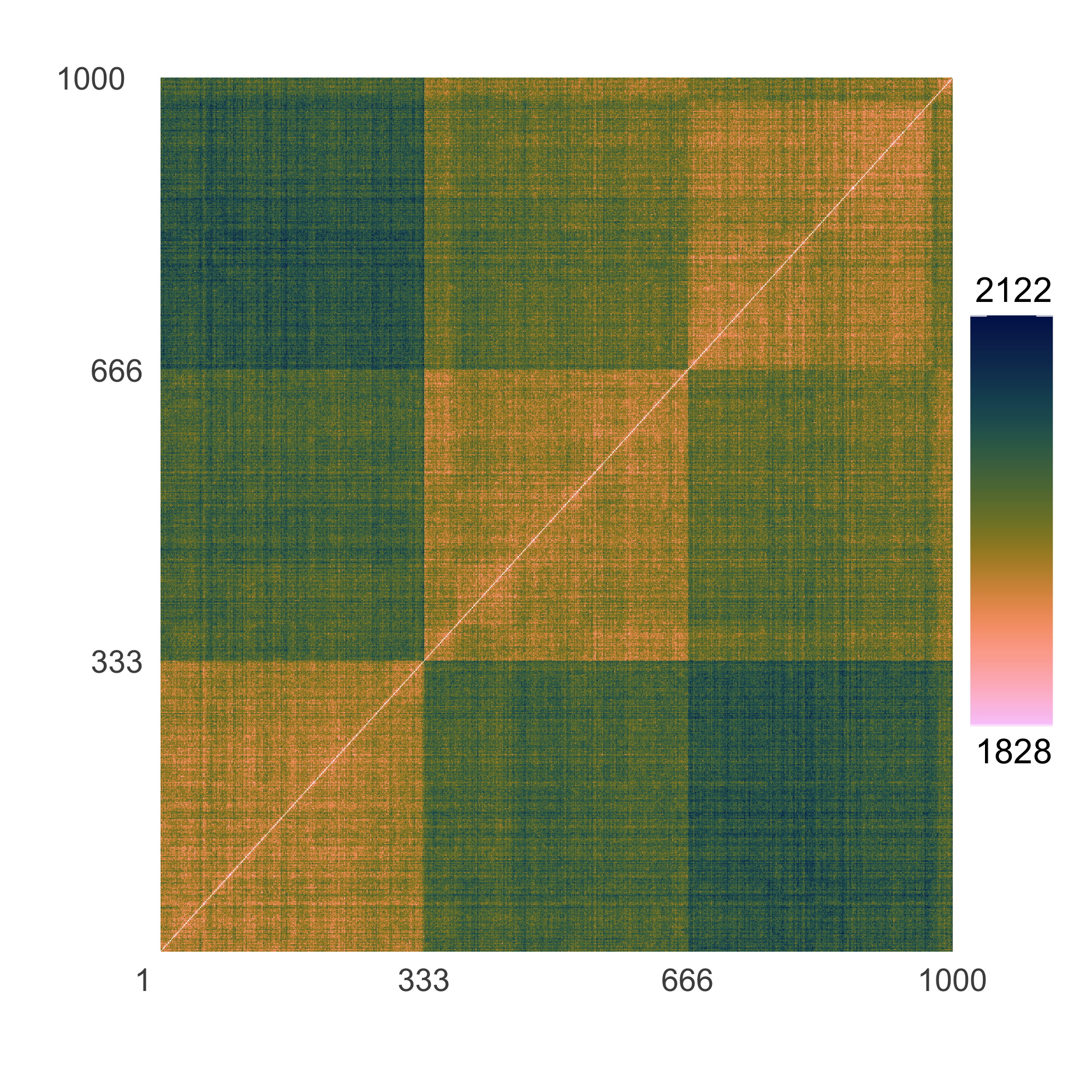}
		\caption{Robinson-Foulds (EBOV trees)}
    \end{subfigure}\hfil
	\begin{subfigure}[b]{2in}
		\includegraphics[scale=0.08]{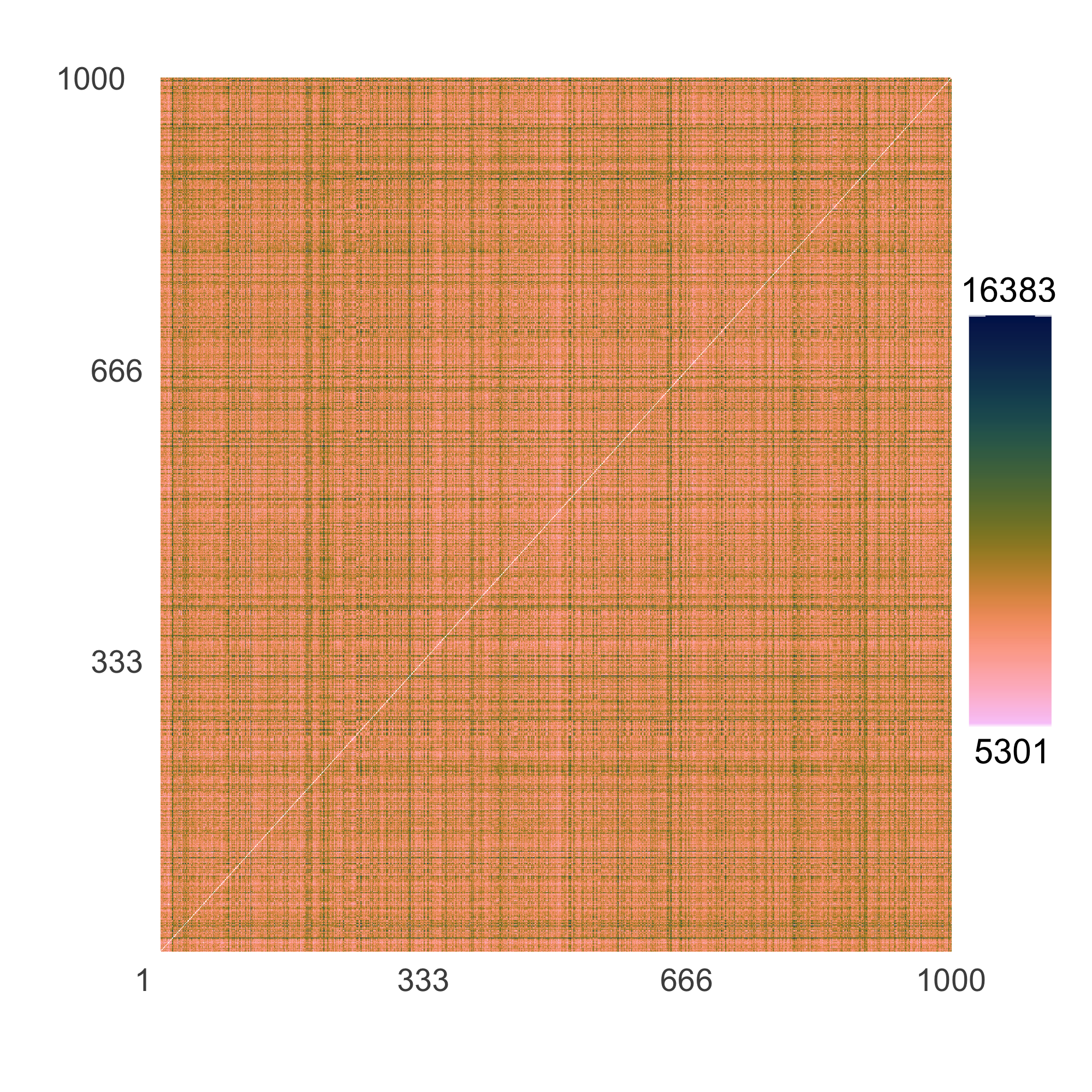}
		\caption{Path difference (EBOV trees)}
    \end{subfigure}\hfil
	\begin{subfigure}[b]{2in}
		\includegraphics[scale=0.08]{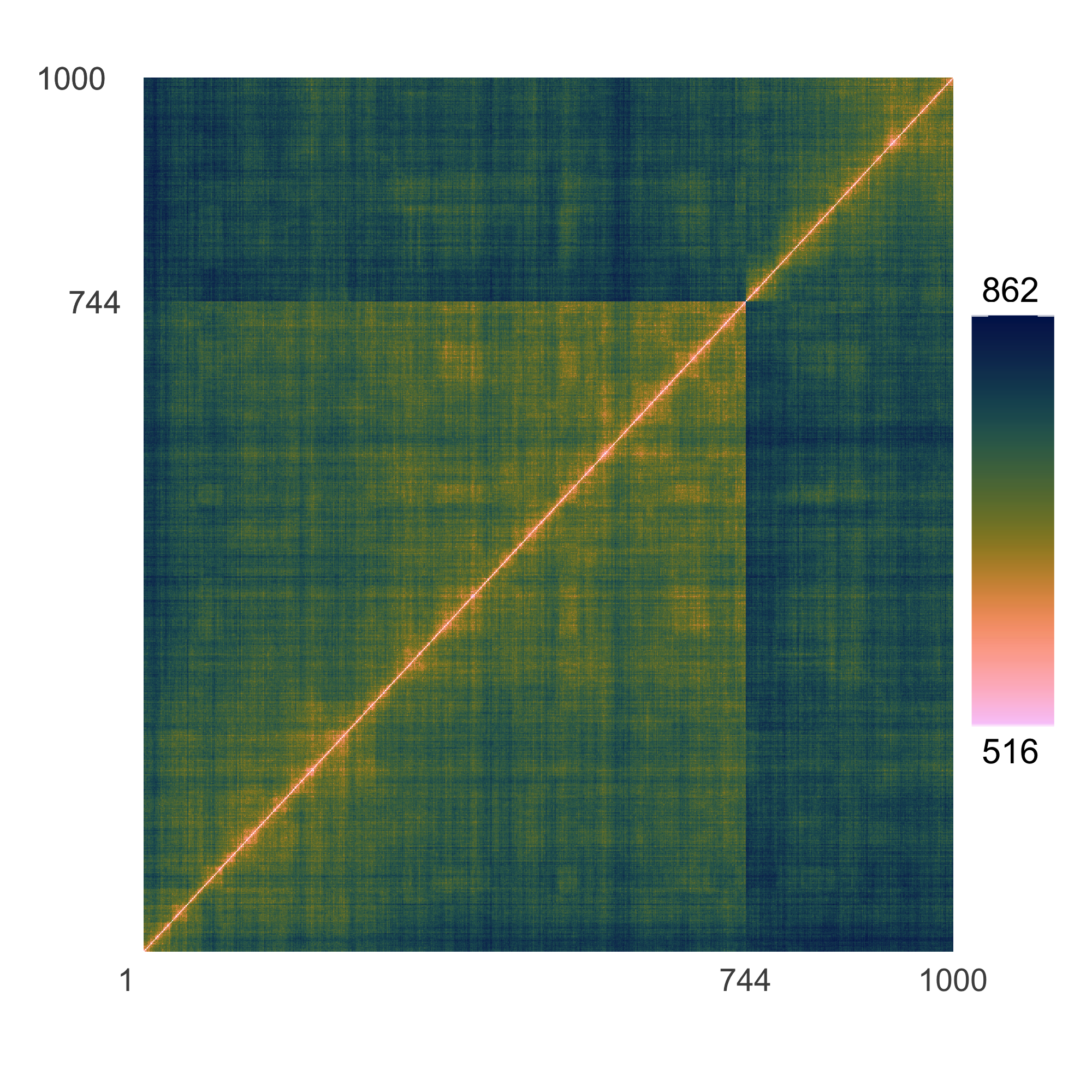}
		\caption{Robinson-Foulds (HIV trees)}
    \end{subfigure}\hfil
	\caption{Heatmaps of pairwise distances between trees in the EBOV and HIV combined sets of samples. For the EBOV trees, the three sets of samples can be clearly distinguished using certain distance metrics (shown here: Robinson-Foulds) but not with others (shown here: path difference). For the HIV trees, the two sets of samples can be distinguished using the Robinson-Foulds distance, which wasn't the case in Figure~\ref{fig:toptracehiv}.}
	\label{fig:heatmaps}
	\end{center}
\end{figure}

A first alternative visualisation comes in the form of heatmaps of the pairwise distances.
Figure~\ref{fig:heatmaps} shows heatmaps of the Robinson-Foulds distances for the EBOV trees and HIV trees, and of the path distances for the EBOV trees.

For the EBOV posterior tree samples, the patterns seen in the trace graphs of Figure~\ref{fig:toptraceebola} are also seen in Figure~\ref{fig:heatmaps}, i.e. distinguishable samples when using the Robinson-Foulds distance, but not when using the path difference.
We refer to Supplementary Figures~\ref{fig:supheatmapsEBOV} and \ref{fig:supheatmapsHIV} for pairwise heatmaps using the other distance metrics, which also recreate the previously observed patterns.

For the HIV posterior tree samples, all heatmaps actually show a clear distinction between the two samples.
This discrepancy -- which was not visible in the trace graphs of Figure~\ref{fig:toptracehiv} -- is thus only produced when considering all pairwise distances as opposed to only distances from the first tree sample.

\subsection*{Network graphs}

\begin{figure}[!h]
	\begin{center}
	\begin{subfigure}[b]{2in}
		\includegraphics[scale=0.35]{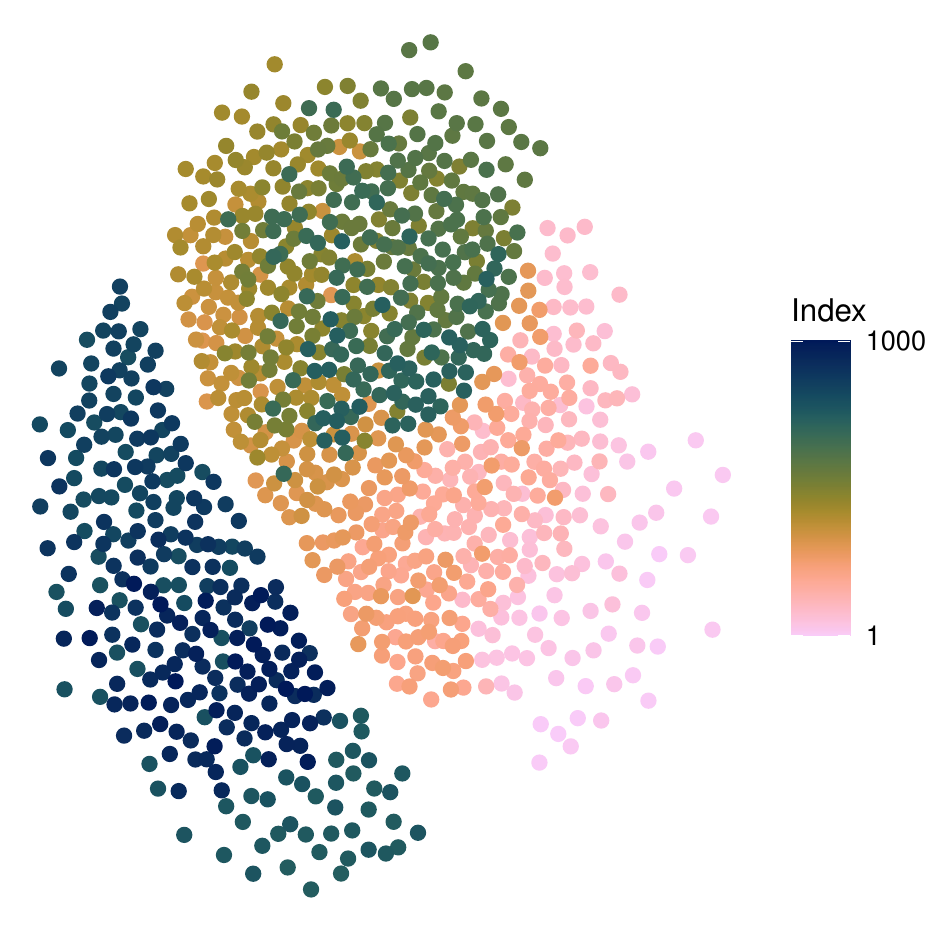}
		\caption{Robinson-Foulds (HIV trees)}
    \end{subfigure}\hfil
	\begin{subfigure}[b]{2in}
		\includegraphics[scale=0.35]{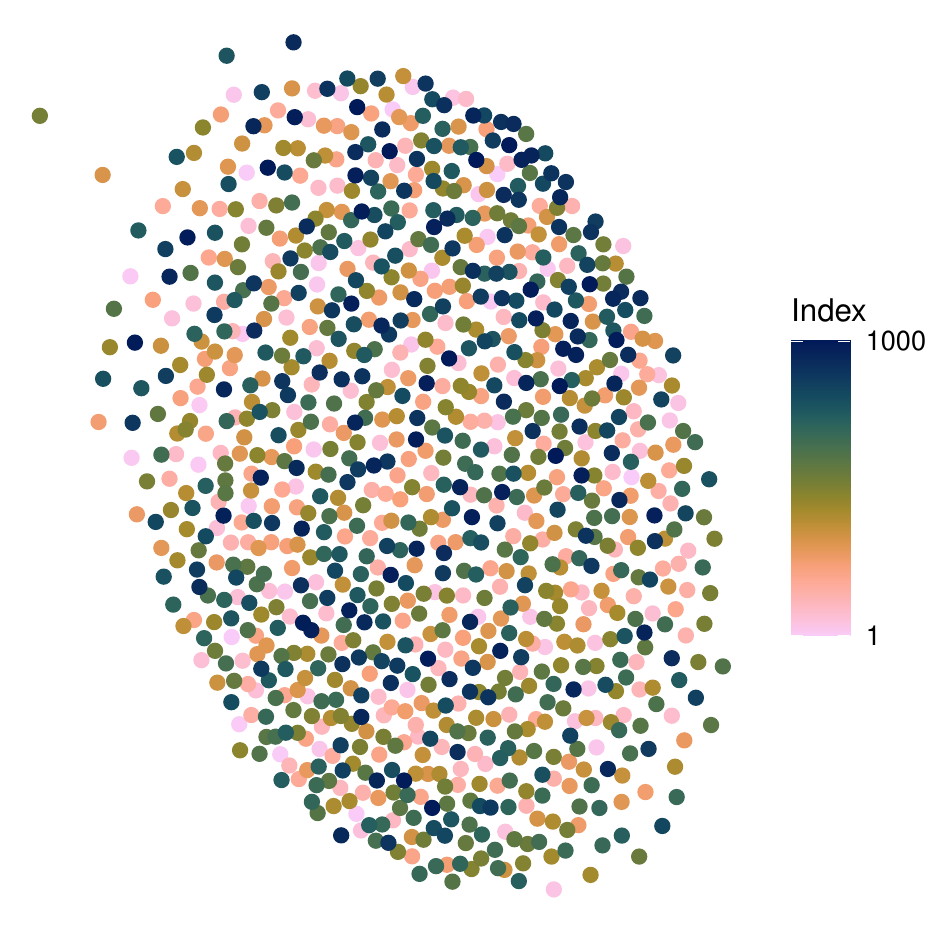}
		\caption{Branch score (HIV trees)}
    \end{subfigure}\hfil
	\begin{subfigure}[b]{2in}
		\includegraphics[scale=0.35]{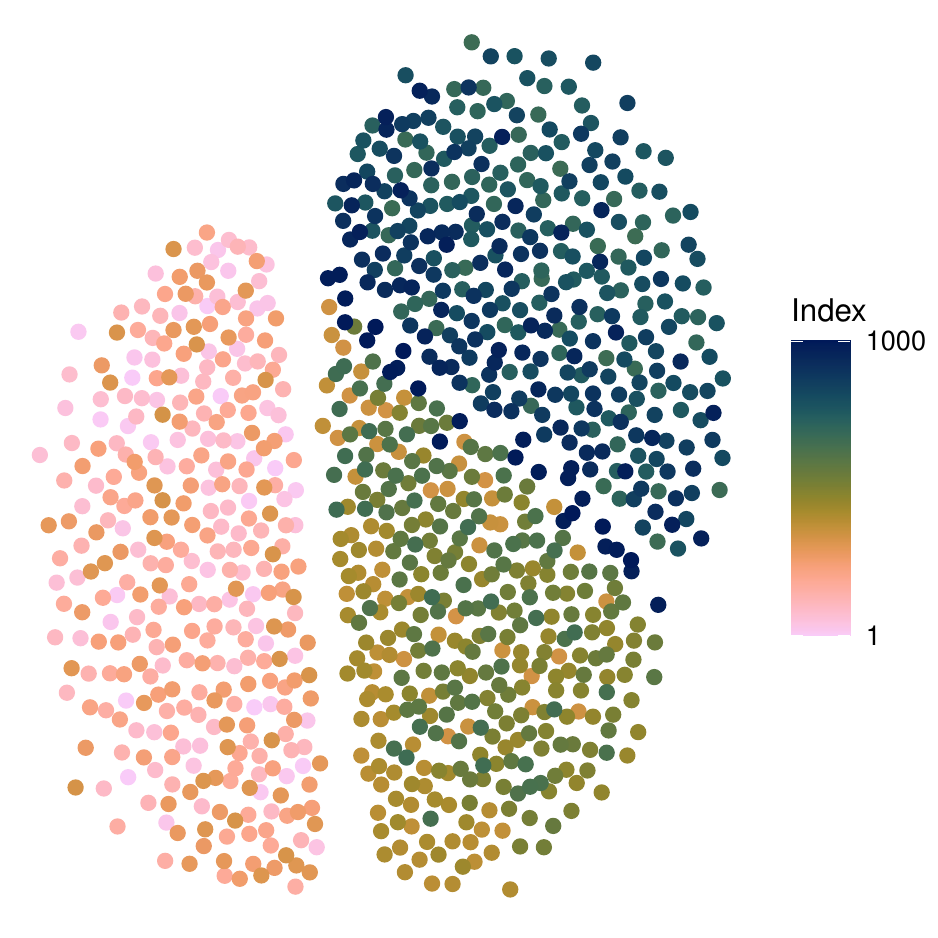}
		\caption{Robinson-Foulds (EBOV trees)}
    \end{subfigure}\hfil
	\caption{Network graphs of pairwise distances between trees in the EBOV and HIV combined sets of samples. Each node reflects a tree, and the relative distances between nodes reflect the pairwise distances between trees. Nodes are coloured by position in the sample, going from light to dark. For the HIV trees, the two sets of samples can be clearly distinguished using the Robinson-Foulds distance, but not using the branch score. For the EBOV trees, the network graph made with the Robinson-Foulds distance shows the three distinct runs clearly.}
	\label{fig:networks}
	\end{center}
\end{figure}

The $1\,000\times1\,000$ matrix of pairwise distances between trees can be converted into a similarity matrix, done here by normalizing the distances to a range of 0 to 1 and substracting them from 1.
Using a force-directed algorithm \citep{fruchterman}, one can create a two-dimensional graph where each tree is represented by a node and relative distances between nodes reflect the pairwise distances between trees.
Figure~\ref{fig:networks} shows the network graphs of the Robinson-Foulds distances for the EBOV and HIV trees, and of the branch score for the HIV trees.
All other network graphs (for both datasets) can be found in Supplementary Figures~\ref{fig:supnnetworksEBOV} and \ref{fig:supnnetworksHIV}.

For the EBOV trees, the patterns we came to expect from the trace graphs and heatmaps are reproduced.
However, for the HIV trees, the two sets of samples can be clearly distinguished using the Robinson-Foulds distance, but not using the branch score.
This is surprising, given how these two metrics are closely related (see Figure~\ref{figure:splitdiffs}).
Furthermore, in the first subsample of the HIV trees (trees 1-744), the colouring of the nodes show a clear gradient---in line with the shape of the trace graphs in Figure~\ref{fig:toptracehiv}.

These visualisations may currently be hard to implement in practice, as the increase in computation time from a simple trace graph (which requires 999 distances in our 1\,000-tree samples) to a set of pairwise distances (499\,500 distances) is substantial---going up to several days on a standard computer.

\section{Non-distance metric based approaches}

All approaches considered in the previous two sections -- with the exception of the split frequency ESS -- explicitly rely on computing phylogenetic distances between trees in a sample. 
We here discuss a few alternative explorations and diagnostics for topological convergence.

\subsection*{Continuous parameters}

\begin{figure}[!h]
	\begin{center}
	\begin{subfigure}[b]{3.3in}
		\includegraphics[width=.95\linewidth]{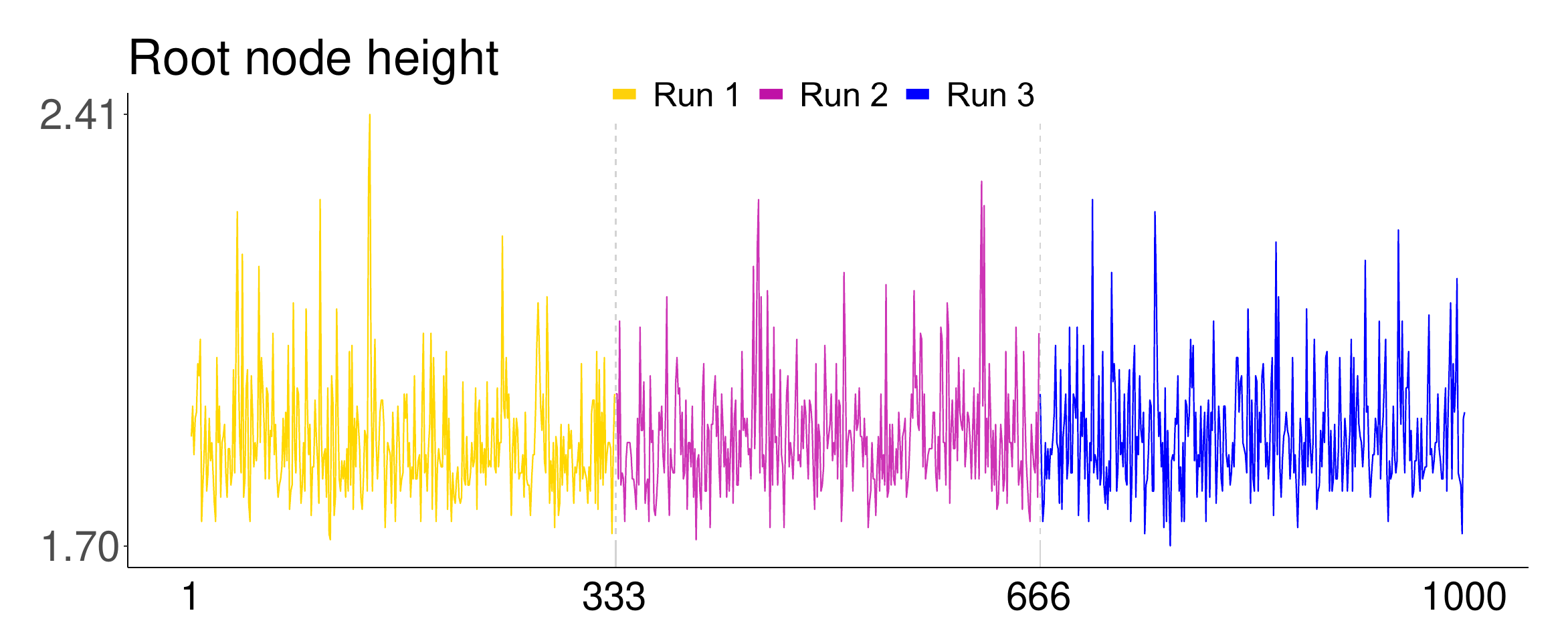}
		\caption{Root node height (EBOV trees) --- $ESS=726$}
    \end{subfigure}\hfil
	\begin{subfigure}[b]{3.3in}
		\includegraphics[width=.95\linewidth]{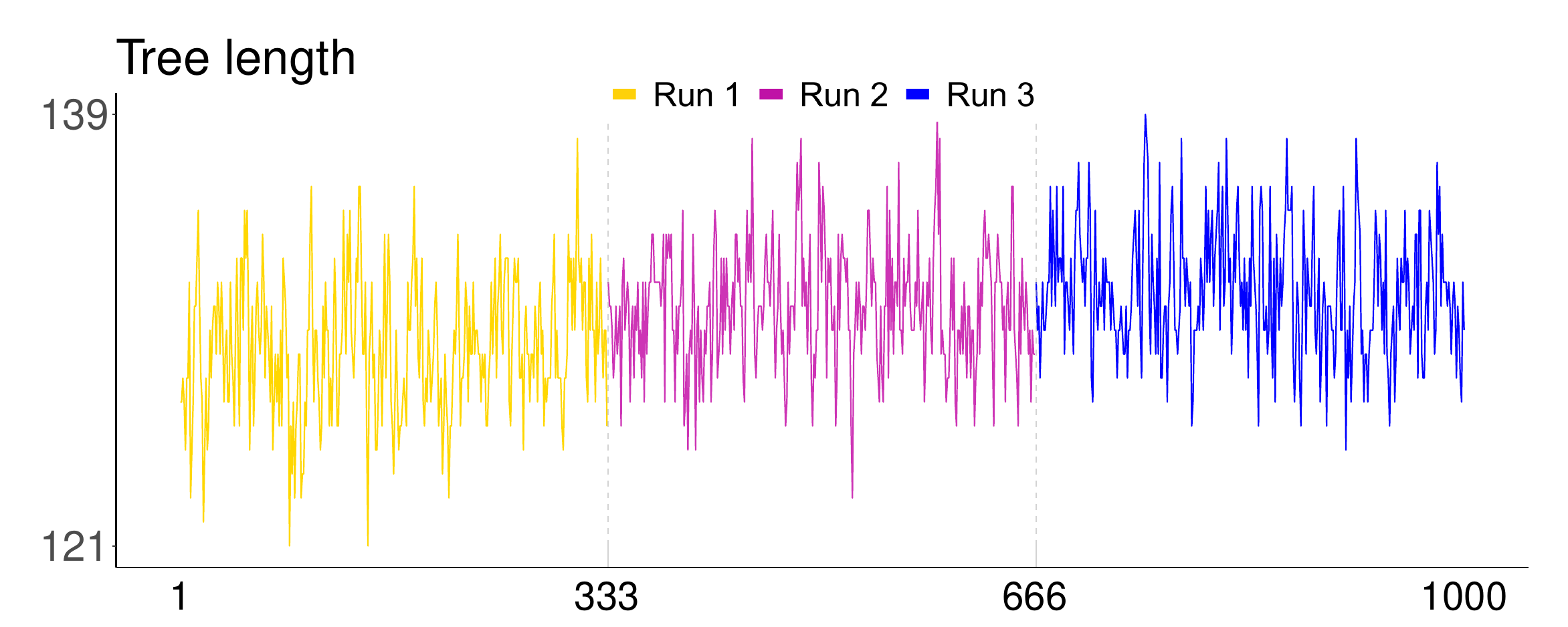}
		\caption{Tree length (EBOV trees) --- $ESS=20$}
    \end{subfigure}
	\caption{Trace plots for the root node height and tree length for the EBOV samples. Tree length refers to the sum of all tree branches, and is thus a statistic closely related to the topology of the tree. The root height, a commonly evaluated parameter, shows satisfactory mixing, while the tree length, which is less often included in convergence / mixing assessment, has an ESS of 20 which indicates problematic mixing. The ESS of individual replicates can be found in tables~\ref{tab:ebovess} and \ref{tab:hivess} for the EBOV and HIV samples respectively.}
	\label{fig:tracerplots}
	\end{center}
\end{figure}

An easy first step would be to consider whether the continuous parameters of the model show any signs of convergence or mixing issues.
While this is standard practice in Bayesian phylogenetic inference, we here pay specific attention to those parameters that directly translate aspects of a phylogenetic tree into continuous values.
Figure~\ref{fig:tracerplots} shows trace graphs of the height of the root node and the total tree length in the EBOV sample.
The height of the root---a continuous parameter often evaluated using standard methodology, as it reflects the temporal distance between the youngest tip and the most recent common ancestor of all sequences---shows a well-mixed sample with a satisfactory ESS.
The trace of the tree length, which is not often readily assessed despite it being readily implemented in the \textit{BEAST} software package, shows three distinguishable subsamples -- although the difference is not marked -- as well as a very poor ESS of only 20.

Supplementary Figure~\ref{fig:suptracerplots} shows the same plots for the HIV sample. In this case, neither of the two statistics show a discrepancy between the two runs and the ESS values are very high.

\subsection*{Maximum clade credibility trees}

The observed discrepancies between individual runs in the explored posterior topological space for both the EBOV and HIV samples raise the question of what these differences actually reflect in terms of phylogenetic inference. 
An often employed summary tree for such samples is the maximum clade credibility (MCC) tree.
We can compute this tree for the total EBOV and HIV samples, as well as the individual subsamples, and compare them.

\begin{figure*}[!htbp]
	\begin{center}
	\includegraphics[scale=0.2]{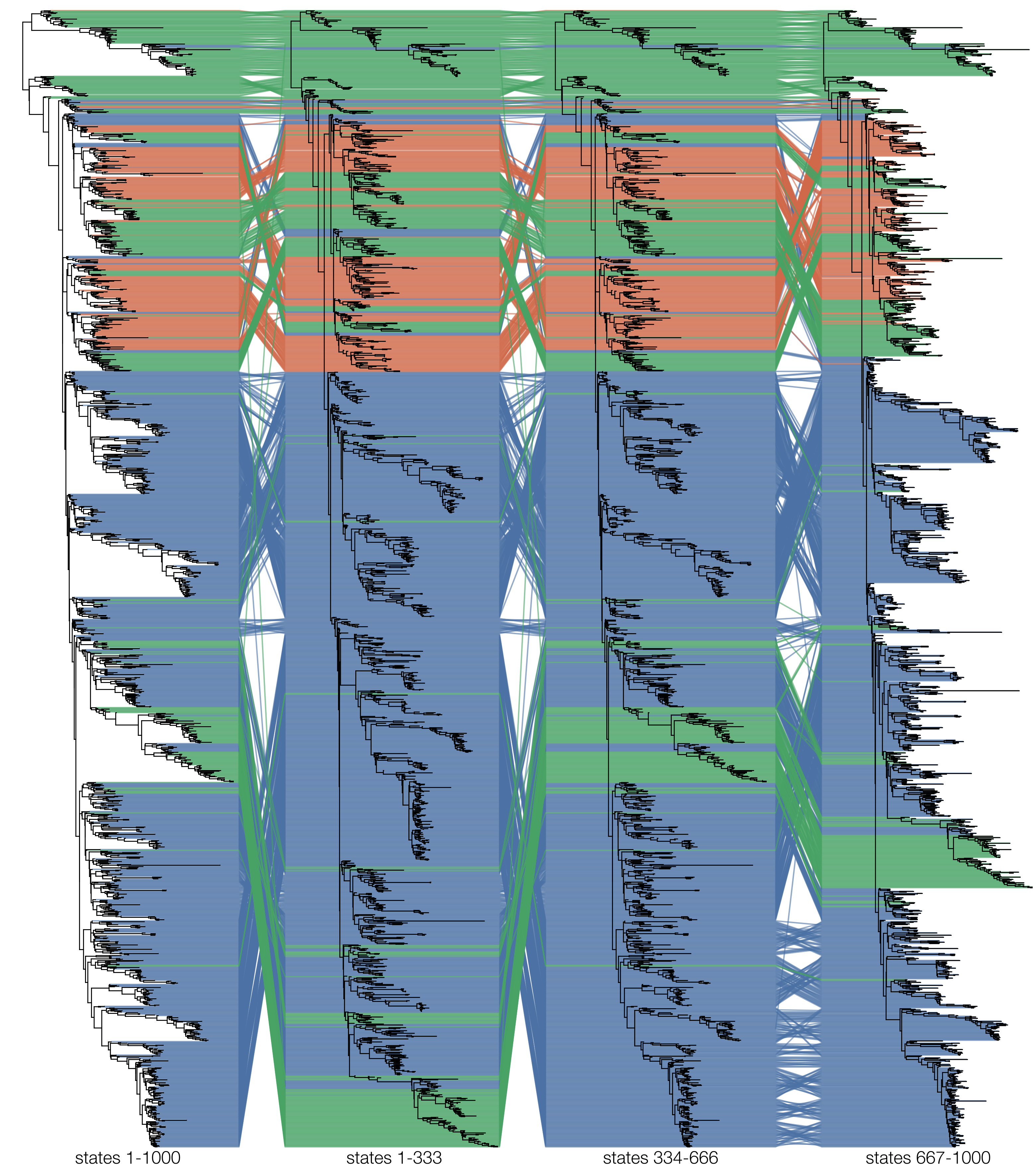}
    \includegraphics[scale=0.4]{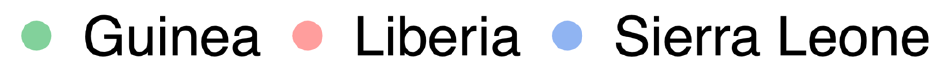}
	\caption{Tanglegram of the MCC trees of the EBOV sample. The tips of the trees are connected to each other by lines, coloured by the country of origin. Disagreement between the subsamples regarding the location of several clades is apparent by the fact that the lines connecting the tips of these clades are not parallel. The MCC tree of the total sample is identical to the MCC tree of the second subsample.} 
	\label{fig:tangleEBOV}
	\end{center}
\end{figure*}

Figure \ref{fig:tangleEBOV} shows a tanglegram comparing the MCC trees of the EBOV samples by linking the corresponding taxa to each other.
Taxa are coloured by country of origin. The trees all have the same overall shape, but several clades end up in different locations depending on the individual run.
Further, Supplementary Figure \ref{fig:tangleEBOVsupp} shows a tanglegram of the MCC trees of the first and second half of the first EBOV run (so trees numbered 1-166 and trees numbered 167-333); these two trees thus reflect within-run variability.
Although these trees have differences as well---as expected from the inherent randomness of the MCMC algorithm---they are substantially more similar than the MCC trees of different replicate analyses, as shown in Figure \ref{fig:tangleEBOV}, thus coinciding with our previous findings that suggested higher between-run variability than within-run variability. 

\begin{figure*}[!htbp]
	\begin{center}
	\includegraphics[scale=0.15]{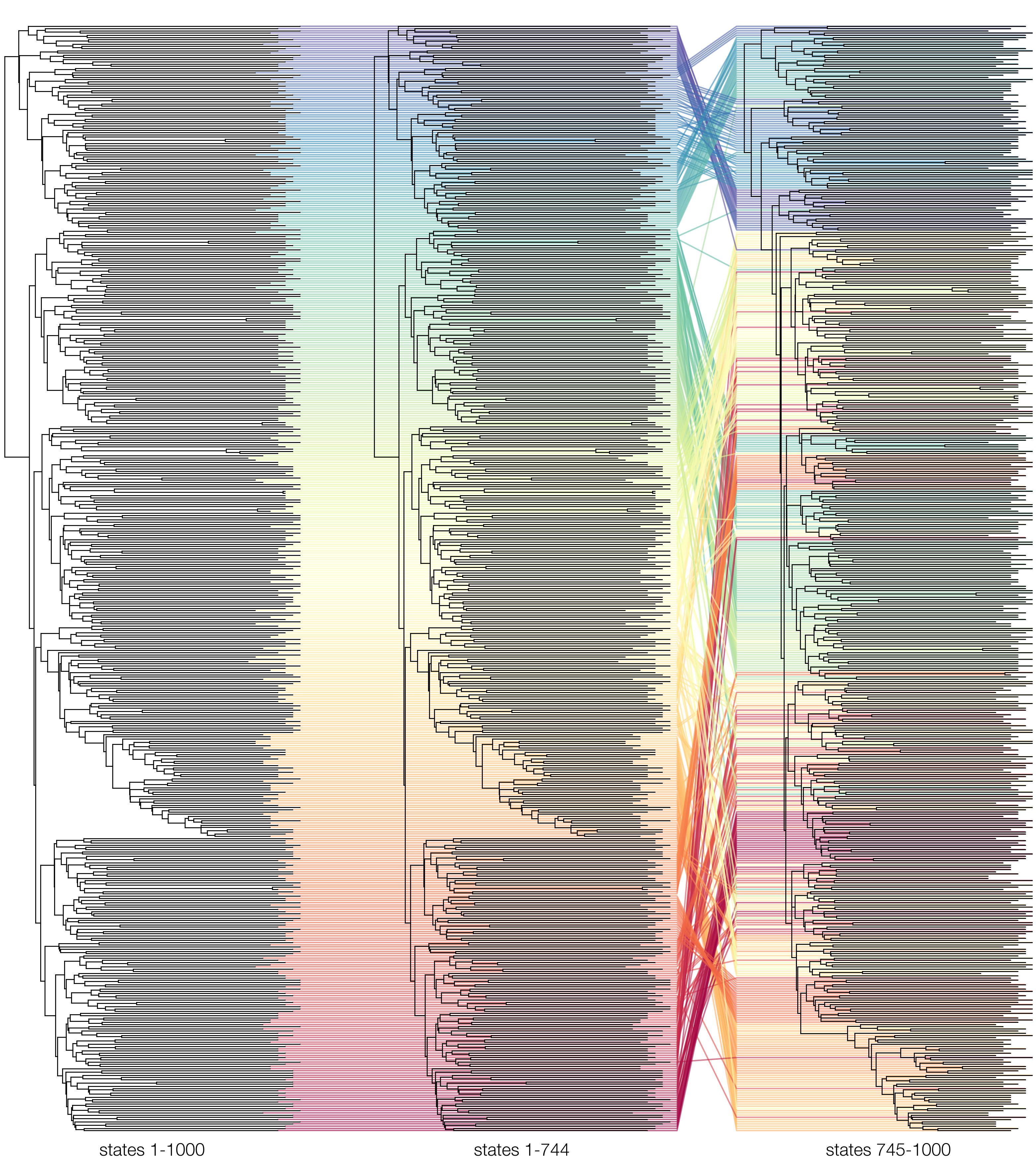}
	\caption{Tanglegram of the MCC trees of the HIV sample. The tips of the trees are connected to each other by lines, coloured by position in the first MCC tree. Disagreement between the subsamples regarding the location of several clades is apparent by the fact that the lines connecting the tips of these clades are not parallel. The MCC tree of the total sample is identical to the MCC tree of the first subsample.}
	\label{fig:tangleHIV}
	\end{center}
\end{figure*}

Similarly, Figure \ref{fig:tangleHIV} shows a tanglegram comparing the MCC trees of the HIV samples by linking the corresponding taxa to each other. 
Taxa are not coloured by geographic origin but simply by location in the sample---as they were in the network graphs of Figure \ref{fig:networks}.
Again, there are substantial differences between the first and second run.
Supplementary Figure \ref{fig:tangleHIVsupp} shows a tanglegram of the MCC trees of the first and second half of the first HIV run (so trees numbered 1-372 and trees numbered 373-744). 
Unlike for the EBOV trees, it is not apparent from the MCC trees that there would be more between-run than within-run variability in the sampled topologies.
This is in line with the low ESS values for the HIV samples -- as shown in Figure~\ref{fig:toptracehiv} -- and can potentially be attributed to the typical star-like shape of HIV phylogenies.

\subsection*{ESS of individual tree splits}

The approach by \citet{Hohna} does not aim to compute an ESS for the entire topology.
Instead, each split in the tree is considered individually as a binary parameter (1=present, 0=absent) for which an ESS is computed using standard methodology.

\begin{figure}[!htbp]
	\begin{center}
	\begin{subfigure}[b]{2.8in}
		\includegraphics[width=.95\linewidth]{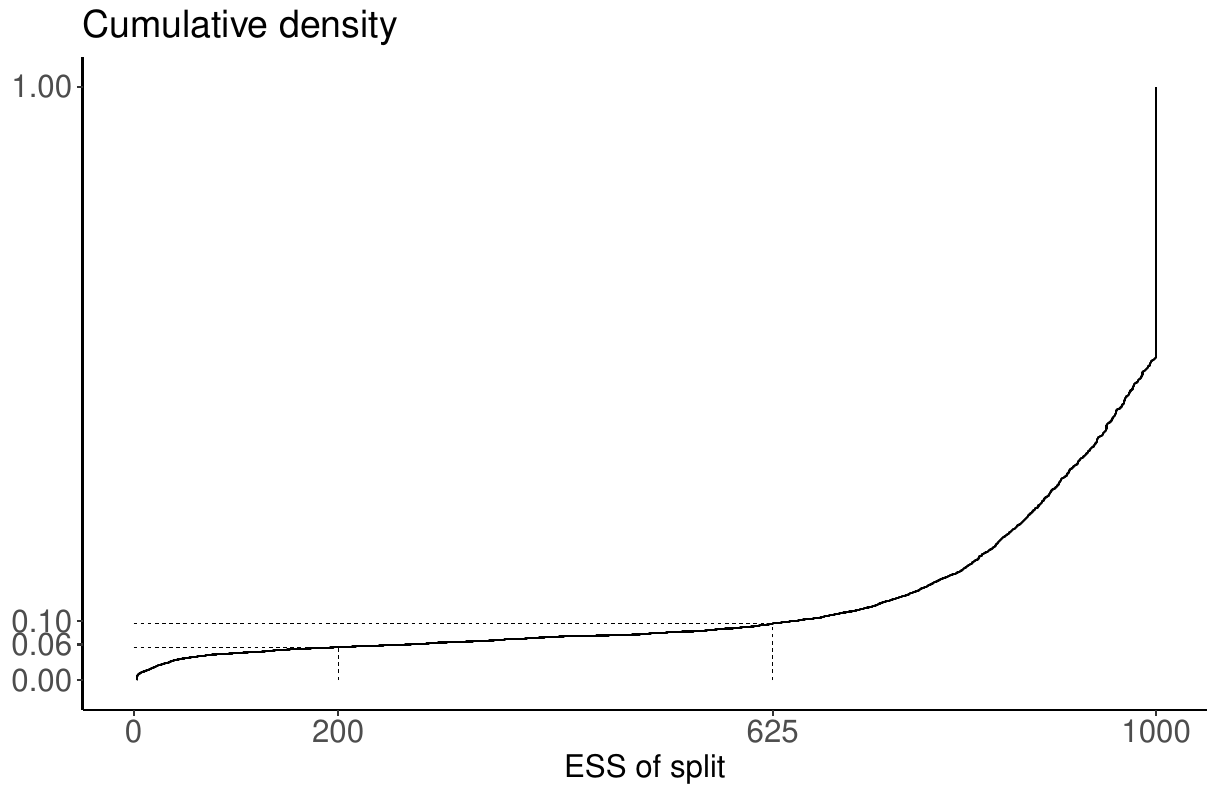}
		\caption{EBOV trees}
    \end{subfigure}\hfil
	\begin{subfigure}[b]{2.8in}
		\includegraphics[width=.95\linewidth]{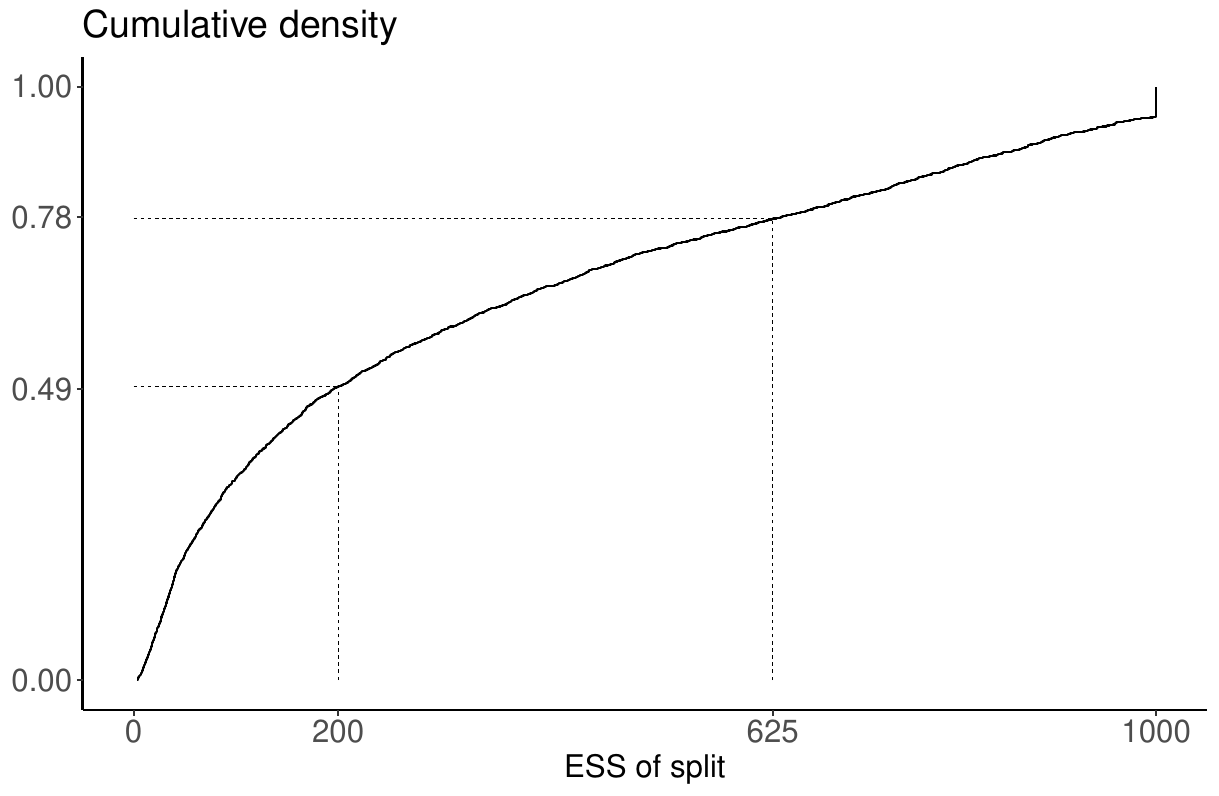}
		\caption{HIV trees}
    \end{subfigure}
	\caption{Cumulative densities of the ESS values of individual splits in the sampled EBOV and HIV trees.
	ESS values of 200 and 625 are indicated and connected to their corresponding cumulative density values with dotted lines. These ESS values were computed using the \texttt{convenience} R package.}
	\label{fig:convenience}
	\end{center}
\end{figure}

Figure~\ref{fig:convenience} shows the cumulative density of ESS values of the $6\,859$ splits observed in the EBOV trees and the $2\,371$ splits observed in the HIV trees.
For the EBOV trees, the vast majority of these (94\%) have an ESS above the often-used cutoff value of $200$. \citet{Hohna} suggest a more stringent cutoff of $625$, which is met by 90\% of the splits.
Although not shown here, the three independent samples that make up the full EBOV sample showed a nearly identically shaped distribution of individual split ESS values.
Thus, the vast majority of splits show a satisfactory ESS both by according to the commonly-used cutoff and more stringent cutoff suggested by \citet{Hohna}.
For the HIV trees, the ESS values are substantially lower, with only 51\% of the splits having an ESS above $200$ and 22\% above $625$.
By both criteria, these ESS values suggest poor mixing of the HIV sample. Thus, when considering the approach towards topological convergence assessment suggested by \citet{Hohna}, the EBOV trees suggest satisfactory results, while the HIV trees do not.


\section{Discussion}

In this case study, we applied an ensemble of diagnostics in topological space on a set of large trees, obtained through Bayesian phylogenetic inference under a set of complex models and through combining the output of at least two independent replicate analyses.
Importantly, we find that the different approaches to evaluating topological convergence can lead to drastically different conclusions, a finding that to the best of our understanding has not been observed to this extent before.
It is likely that a combination of the complexity of the models, the size of the datasets, and the fact that we looked at samples that result from two or three different / independent analysis replicates---as opposed to only just one---allowed us to observe these phenomena that went unnoticed before.
These findings stress the importance of assessing topological convergence to the posterior and not merely continuous parameter and (joint) density convergence, which is the current approach in nearly all Bayesian phylogenetic and phylodynamic studies.
Even when additional continuous parameters -- such as the root height and the tree length -- are to be logged for assessing their ESS, these still offer no guarantee at avoiding topological convergence difficulties.
Whether the discrepancies we found affect inferences on estimates of parameters of interest downstream in the analysis is not yet clear, and warrants further research.

\subsection*{The importance of performing replicate analyses}

We emphasize the importance of running more than one replicate analysis---using different starting points---when performing Bayesian phylodynamic inference, as it is clear from our results that even a well-behaved sample from a single replicate may not be representative of the posterior topological space. 
When performing multiple replicate / independent analyses, it's important to favor a few long runs over many short runs (see section 1.11.3 in \citet{brooks2011handbook}), as many short runs can keep one from running the analysis long enough to detect pseudo-convergence (i.e., when the Markov chain appears to have converged but not to the true posterior distribution, possibly due to parts of the state space being poorly connected by the Markov chain dynamics which means that it takes many iterations to get from one part to another) or other problems.

\subsection*{Visualising topological convergence}

\begin{figure}[!h]
	\begin{center}
	\begin{subfigure}[b]{3.3in}
		\includegraphics[width=.95\linewidth]{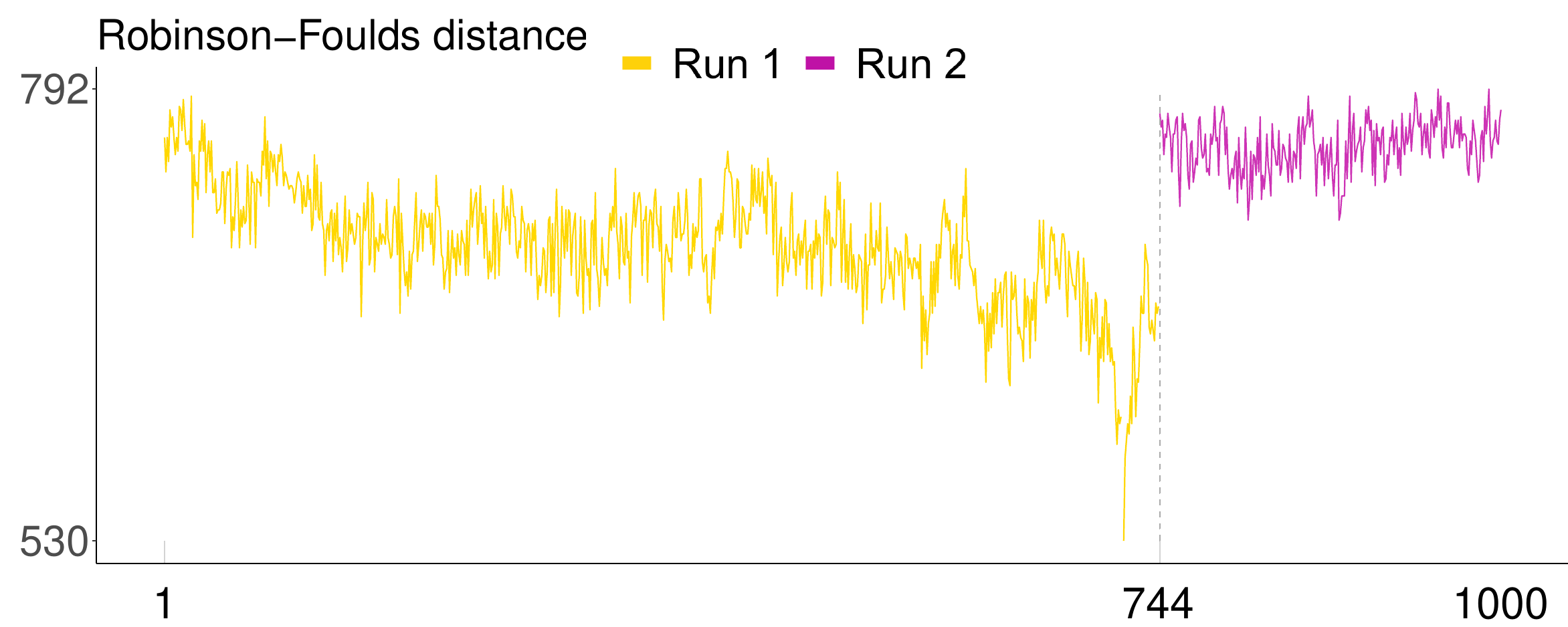}
		\caption{Distance from tree 717 (HIV trees)}
    \end{subfigure}\hfil
	\begin{subfigure}[b]{3.3in}
		\includegraphics[width=.95\linewidth]{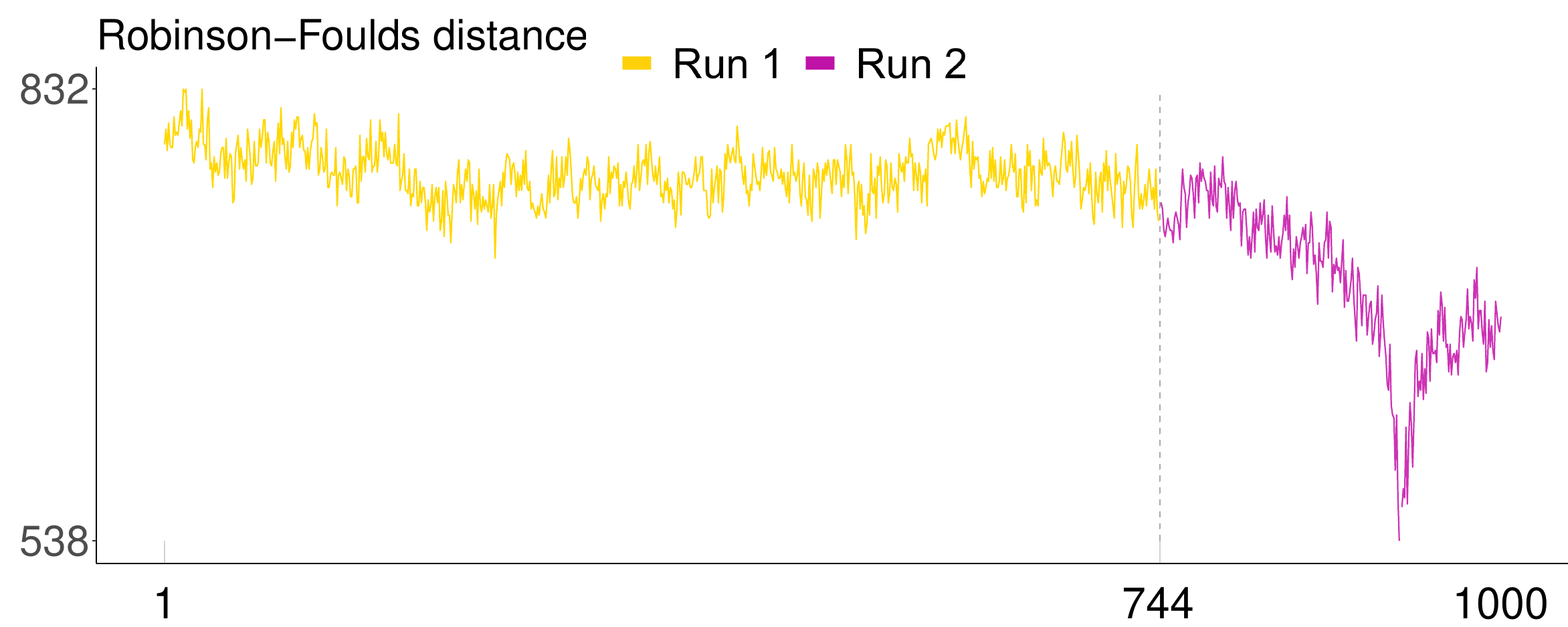}
		\caption{Distance from tree 925 (HIV trees)}
    \end{subfigure}\hfil
	\caption{Trace plots of the Robinson-Foulds distance to each tree using the MCC trees of the first and second subsample (trees 717 and 925 respectively) as the reference tree for the HIV analysis.}
	\label{fig:traceplotmcc}
	\end{center}
\end{figure}

The most straightforward approach to visualising topological convergence is the topological trace plot. 
Making such a graph requires the choice of a reference tree, which can have a substantial impact on the ability of the trace to discriminate between runs.
For example, the two HIV runs could not be distinguished in any of the topology trace graphs in Figure~\ref{fig:toptracehiv}, which used the first tree in the sample as a reference tree from which to compute distances---but were clearly distinct when considering all pairwise distances in Figures~\ref{fig:heatmaps} and \ref{fig:networks}.
Figure~\ref{fig:traceplotmcc} shows the trace plots of the Robinson-Foulds distance to each tree using the MCC trees of the first and second subsample as the reference trees, instead of simply the first.
The discrepancy between individual runs is much more apparent here than in the trace plots of Figure~\ref{fig:toptracehiv}.
This suggests that the choice of reference tree can have a substantial impact on the ability of topological traces to discriminate between runs, and is an important consideration when interpreting these diagnostics.
Furthermore, the trace graphs of Figure~\ref{fig:toptracehiv} had a trajectory reminiscent of undiscarded burn-in. However, the trace graphs of Figure~\ref{fig:traceplotmcc} show a pronounced slump around the reference (MCC) trees, which cannot be explained by burn-in. 
A possible explanation would be that---despite thinning of the chain and there being no sign of any problems with the transition kernels---the chain could exhibit very strong autocorrelation leading to the trees closer to the reference tree simply being more similar to it. 
Given that the reference tree in Figure~\ref{fig:toptracehiv} was the first tree, it can be difficult to tell whether the slump is due to autocorrelation or burn-in.

The issue of selecting a reference tree is resolved by considering pairwise distance based visualisations such as heat maps and network graphs instead of a topological trace plot, making such visualisations preferable. 
However, the computational cost of calculating $n\times(n-1)/2$ distances for a sample of $n$ trees quickly ramps up as $n$ increases. A possible solution would be to reduce the sample size for the purposes of convergence assessment by only using a subset of equally spaced trees---such as was done in Supplementary Figures~\ref{fig:supnnetworksEBOV} (e) and ~\ref{fig:supnnetworksHIV} (e). This can substantially reduce the computational requirements, at the cost of not considering every tree in the sample.

\subsection*{The choice of distance metric}

The impact of the choice of topological distance metric on the ability to detect discrepancies in the combined chains cannot be systematically determined in a case-study format such as this.
Comparing trees using SPR distances---which are closely related to how the studied trees were generated since most implemented MCMC transition kernels use SPR-like moves for exploring tree-space---did suggest that the trees explored by the replicate analyses differ in a topologically substantial manner.
In the EBOV data, the metrics which detected differences between replicates were those based on splits in the trees, while the metrics that could not were those based on tip-to-tip distance.  
In the HIV data however, the difference in performance between these two categories was much less clear, with metrics from both groups being both capable and incapable of detecting differences between replicates. 
We speculate that tree shape---with the HIV data in this analysis being characteristically ``star-like'' with longer branches towards the tips, and the EBOV data being more “ladder-like” \citep{treeshape}---may have an impact on how sensitive different topological distance metrics are to changes in topology. 

While we would recommend using a variety of distance metrics when performing topological convergence assessment, the unweighted Robinson-Foulds distance and the SPR distance are good starting points given that they were the only ones capable of capturing the discrepancies between the independent replicates for both the EBOV and HIV data.
As for which ESS estimator to use, it is difficult to make strong recommendations based on our current results, given that the behaviour of the different estimators depended both on the dataset and the distance metric used.
We would restate the conclusions made by \citet{Magee} that the pseudo-ESS and the Fréchet correlation ESS are the optimal choice according to their experiments, although we recommend looking at the entire range of pseudo-ESS values (not just the median and minimum), since the choice of reference tree can have a large impact on its value.

\subsection*{On combining the output of independent replicates}

Finally, our findings raise important questions as to how the output of replicate Bayesian phylogenetic and phylodynamic analyses should be combined when discrepancies in topological space are detected.
The EBOV and HIV replicates produced samples from different regions of the posterior distribution, and the sampled trees from these replicates were systematically more different than trees from the same replicate, which suggests that these replicates were stuck in local modes of the topological space.
A key open question stemming from our work is how to combine these sampled trees in such a way that the resulting summary tree accurately reflects the region of highest posterior topological probability as well as the uncertainty surrounding it.

Simply averaging the samples with weights proportional to their size---which is currently standard practice---might not produce estimates that properly reflect the multimodal posterior.
As with the HIV example, where the two subsamples were of unequal size, the region explored by the first subsample represents almost three fourths of the total sample, a degree of representation that is unlikely proportional to the posterior mass of this region compared to the second subsample.
From Figure~\ref{fig:tangleHIV}, the MCC tree of the total HIV sample is the MCC tree of the first subsample. 
However, when we artificially lengthen the second subsample by duplicating it two additional times, such that the first and second subsample are roughly of equal length, we find that the MCC tree of the total sample is no longer the initial estimate. 
Thus, if concatenated samples are exploring different regions of topological space, their relative weights in the total sample have a meaningful impact on downstream inferences.
Techniques such as importance sampling could be employed to weigh samples proportionally to their posterior density \citep{yao2022stacking}, such that relative representation of different regions of the posterior are preserved, but we considered these to deserve additional attention and out of scope of the current manuscript.

\section{Competing interests}
No competing interest is declared.

\section{Author contributions statement}
M.B. and G.B. initialised the study.
M.B. performed the analyses and wrote the manuscript. 
G.D. created the tanglegrams and computed the aSPR distances. 
L.M.C., J.G., F.A.M., A.R., M.A.S., G.D., S.L., P.L., and G.B. provided valuable guidance and feedback. 
All authors reviewed the manuscript.

\section{Acknowledgments}

The authors thank Barney Isaksen Potter for their help developing quality figures and Andrew Magee for their valuable technical support.
S.L.H. and G.B. acknowledge support from the Research Foundation - Flanders (``Fonds voor Wetenschappelijk Onderzoek - Vlaanderen,'' G0E1420N).
G.B. acknowledges support from the Internal Funds KU Leuven (Grant No. C14/18/094), from the Research Foundation - Flanders (``Fonds voor Wetenschappelijk Onderzoek - Vlaanderen,'' G098321N) and from the European Union Horizon 2023 RIA project LEAPS (grant agreement no. 101094685).
P.L., M.A.S. and A.R. acknowledge support from the Wellcome Trust (Collaborators Award 206298/Z/17/Z, ARTIC network), the European Research Council (grant agreement no. 725422 -- ReservoirDOCS) and the National Institutes of Health (NIH) (R01 AI153044).
F.A.M. and M.A.S. acknowledge further support from the NIH through R01 AI162611.
P.L. acknowledges support from the Research Foundation, Flanders (``Fonds voor Wetenschappelijk Onderzoek - Vlaanderen,'' G066215N, G0D5117N and G0B9317N) and from the European Union Horizon 2020 project MOOD (grant agreement no. 874850).
M.B. and G.B. acknowledge support from the DURABLE EU4Health project 02/2023-01/2027, which is co-funded by the European Union (call EU4H-2021-PJ4) under Grant Agreement No. 101102733.
Views and opinions expressed are however those of the author(s) only and do not necessarily reflect those of the European Union or the European Health and Digital Executive Agency.
Neither the European Union nor the granting authority can be held responsible for them.

\begin{figure}[H]
    \centering
    \includegraphics[width=0.25\textwidth]{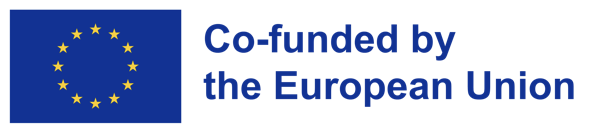}
\end{figure}

\bibliographystyle{abbrvnat}
\bibliography{reflection.bib}

\clearpage
\newpage


\setcounter{figure}{0}

\makeatletter 
\renewcommand{\thefigure}{S\@arabic\c@figure}
\renewcommand{\thetable}{S\@arabic\c@table}
\makeatother

\section{Supplementary Materials}

\begin{figure*}[h]
	\begin{center}
		\begin{subfigure}[t]{1.7in}
			\includegraphics[scale=0.3]{figures/NetworkEBOV_RF.pdf}
		\caption{Robinson-Foulds}
    \end{subfigure}\hfil
	\begin{subfigure}[t]{1.7in}
		\includegraphics[scale=0.3]{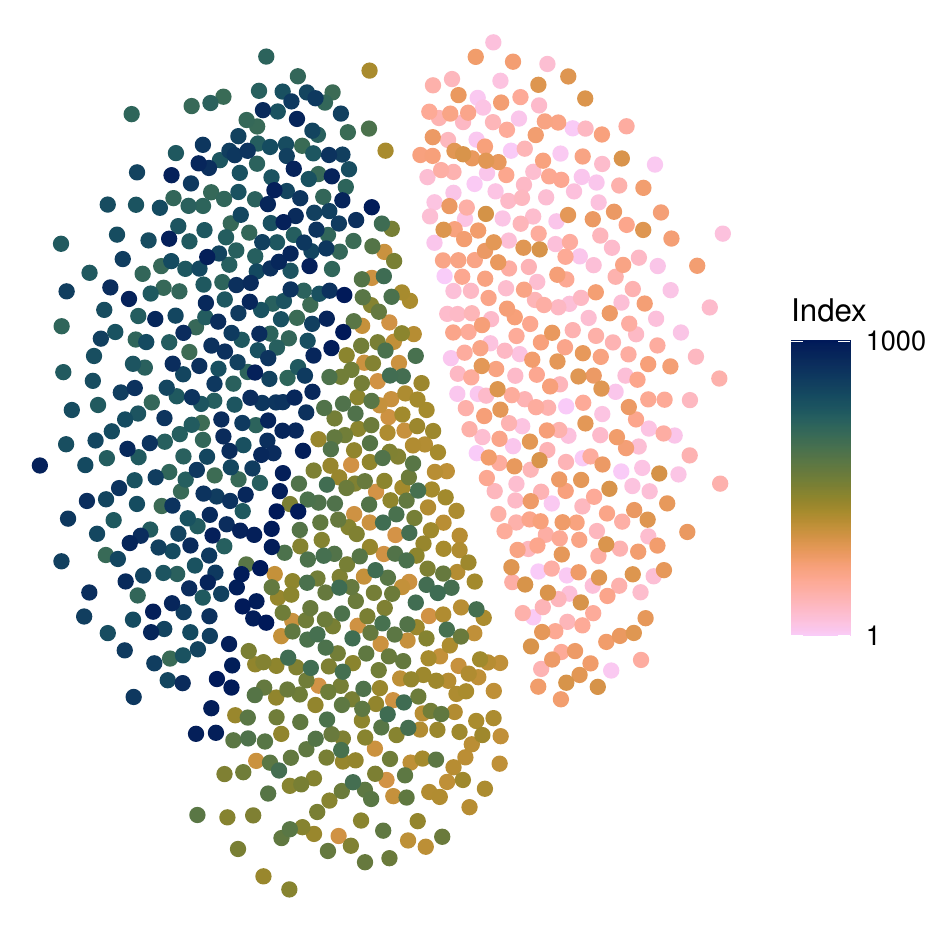}
		\caption{Weighted Robinson-Foulds}
    \end{subfigure}\hfil
	\begin{subfigure}[t]{1.7in}
		\includegraphics[scale=0.3]{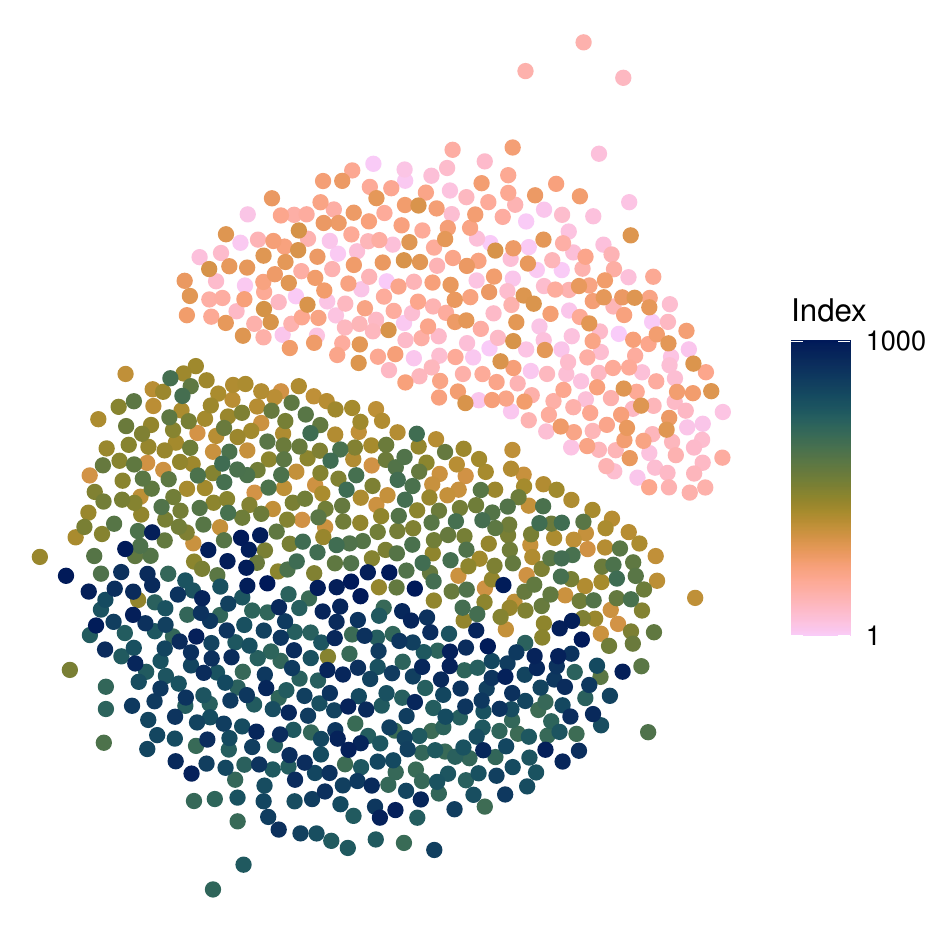}
		\caption{Branch score}
    \end{subfigure}\hfil
	\begin{subfigure}[t]{1.7in}
		\includegraphics[scale=0.3]{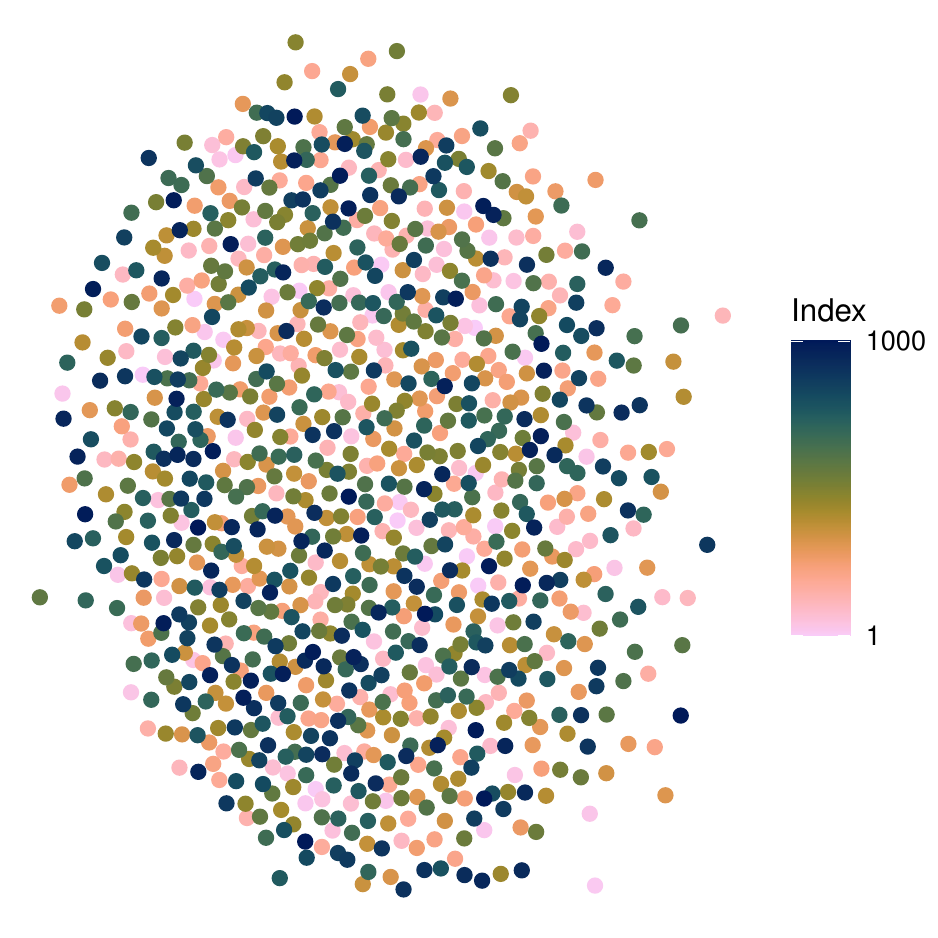}
		\caption{Path difference}
    \end{subfigure}\hfil
	\begin{subfigure}[t]{1.7in}
		\includegraphics[scale=0.3]{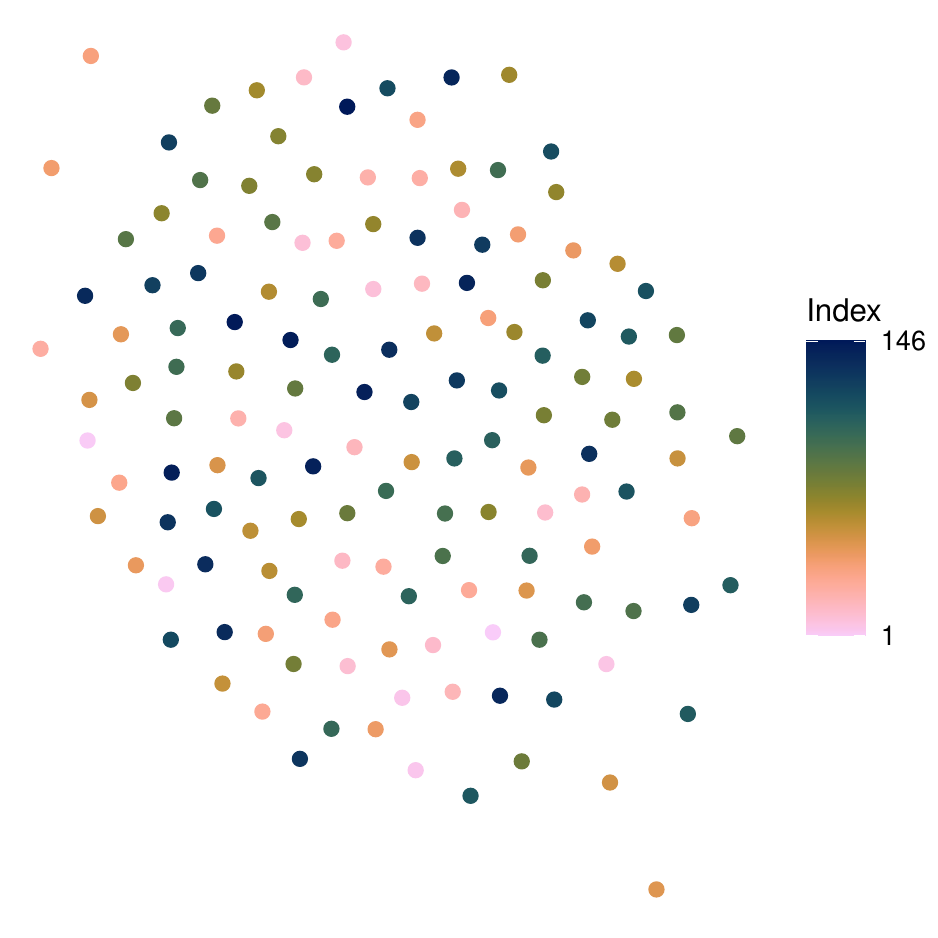}
		\caption{Kendall-Colijn ($\lambda=0$)}
    \end{subfigure}\hfil
	\begin{subfigure}[t]{1.7in}
		\includegraphics[scale=0.3]{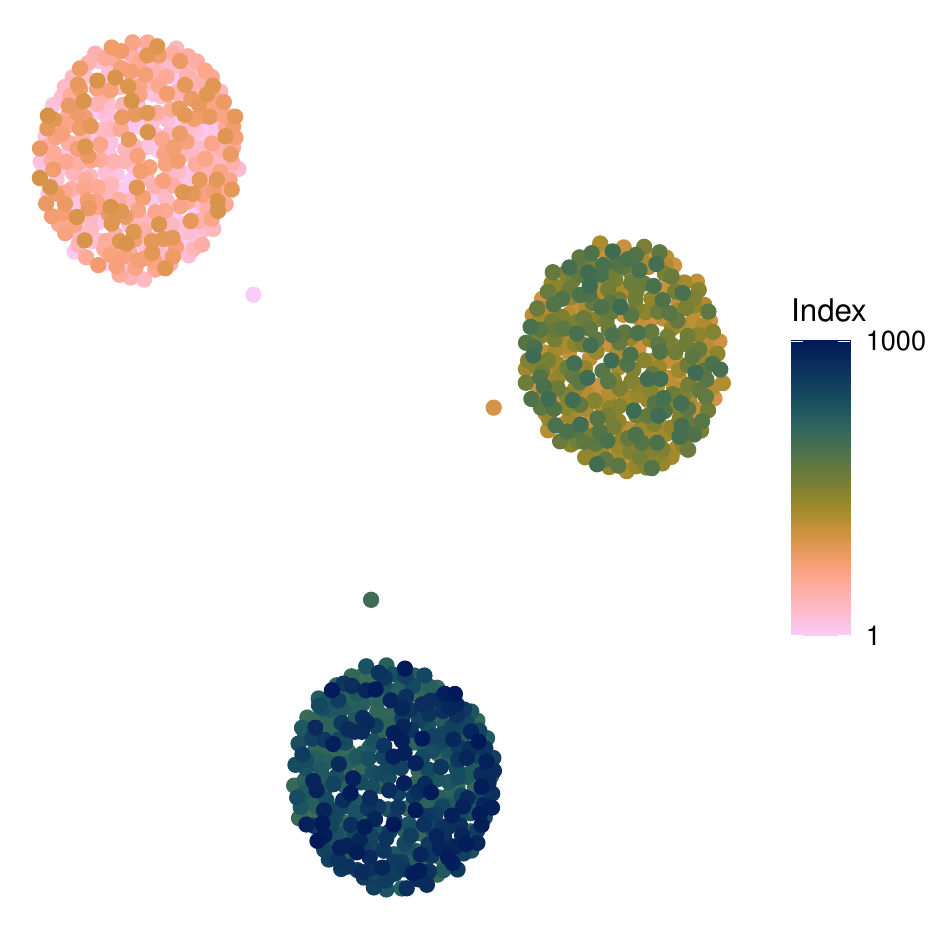}
		\caption{aSPR}
    \end{subfigure}\hfil
	\caption{EBOV networks for all phylogenetic distance metrics. For the Kendall-Colijn distance, the full sample was downsampled to 146 trees equally spaced between tree 1 and tree 1000, as computing pairwise distances for 1000 trees was not feasible.}
	\label{fig:supnnetworksEBOV}
	\end{center}
\end{figure*}

\begin{figure*}[h]
	\begin{center}
	\begin{subfigure}[t]{1.8in}
		\includegraphics[scale=0.06]{figures/HeatmapEBOV_RF.png}
		\caption{Robinson-Foulds}
    \end{subfigure}\hfil
	\begin{subfigure}[t]{1.8in}
		\includegraphics[scale=0.06]{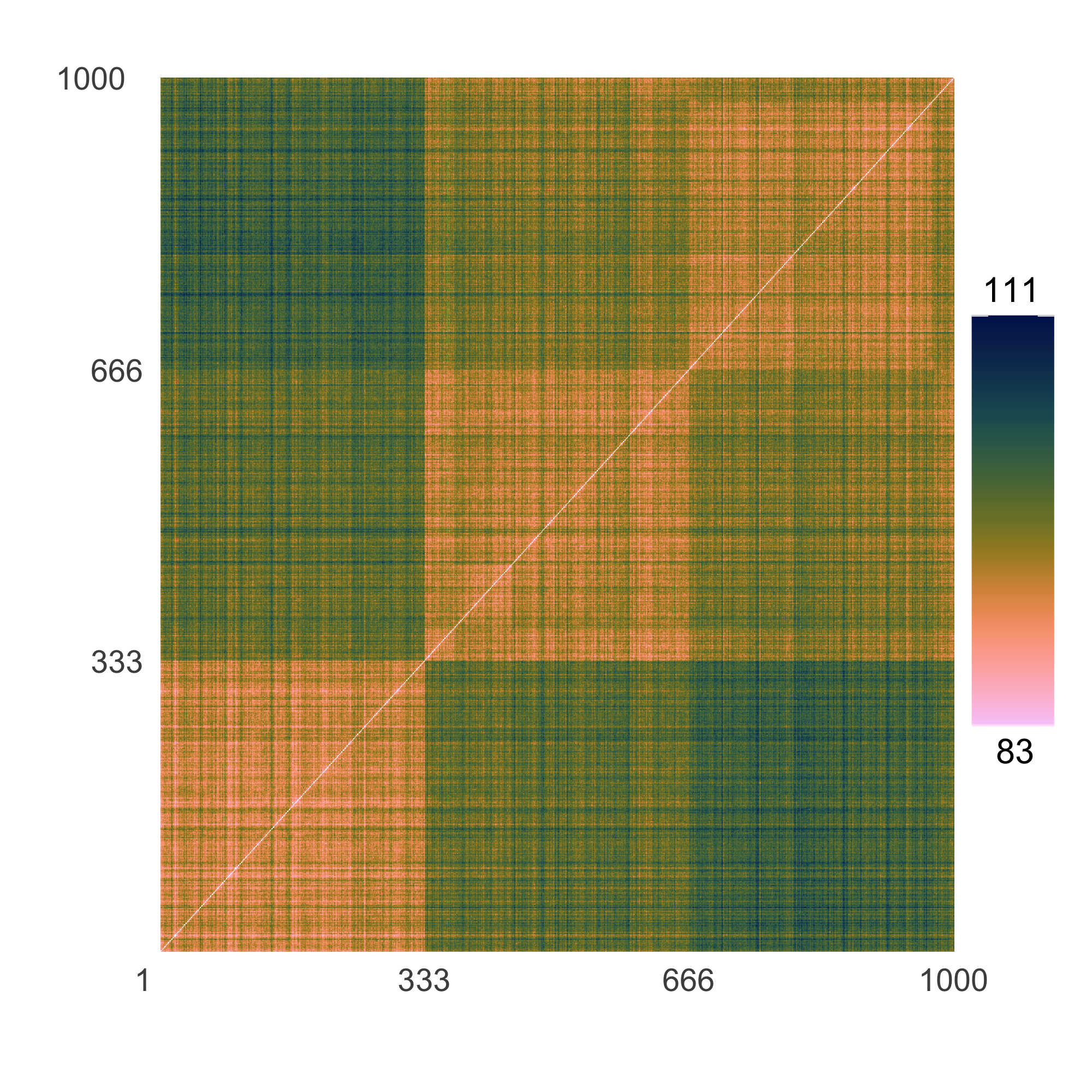}
		\caption{Weighted Robinson-Foulds}
    \end{subfigure}\hfil
	\begin{subfigure}[t]{1.8in}
		\includegraphics[scale=0.06]{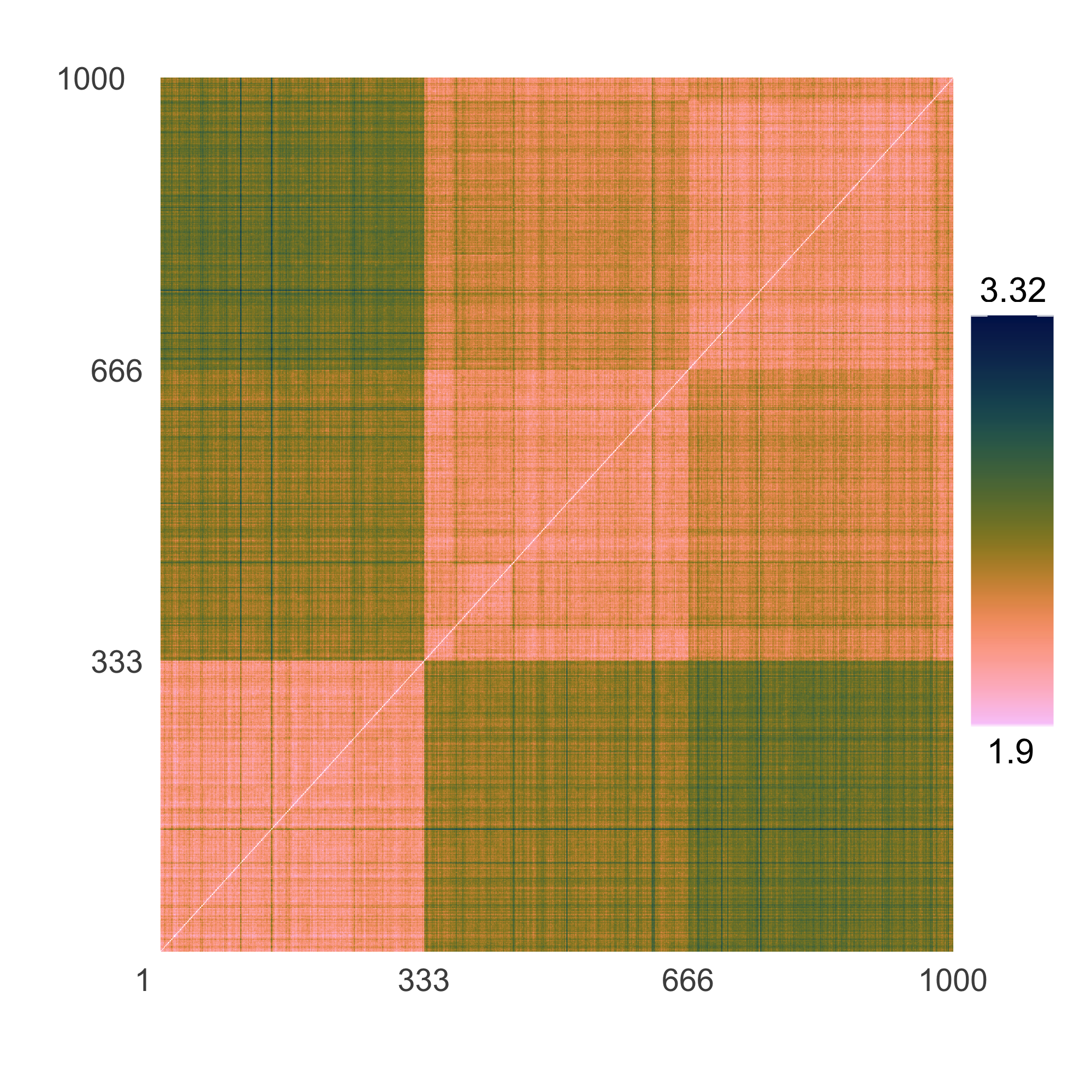}
		\caption{Branch score}
    \end{subfigure}\hfil
	\begin{subfigure}[t]{1.8in}
		\includegraphics[scale=0.06]{figures/HeatmapEBOV_PD.png}
		\caption{Path difference}
    \end{subfigure}\hfil
	\begin{subfigure}[t]{1.8in}
		\includegraphics[scale=0.06]{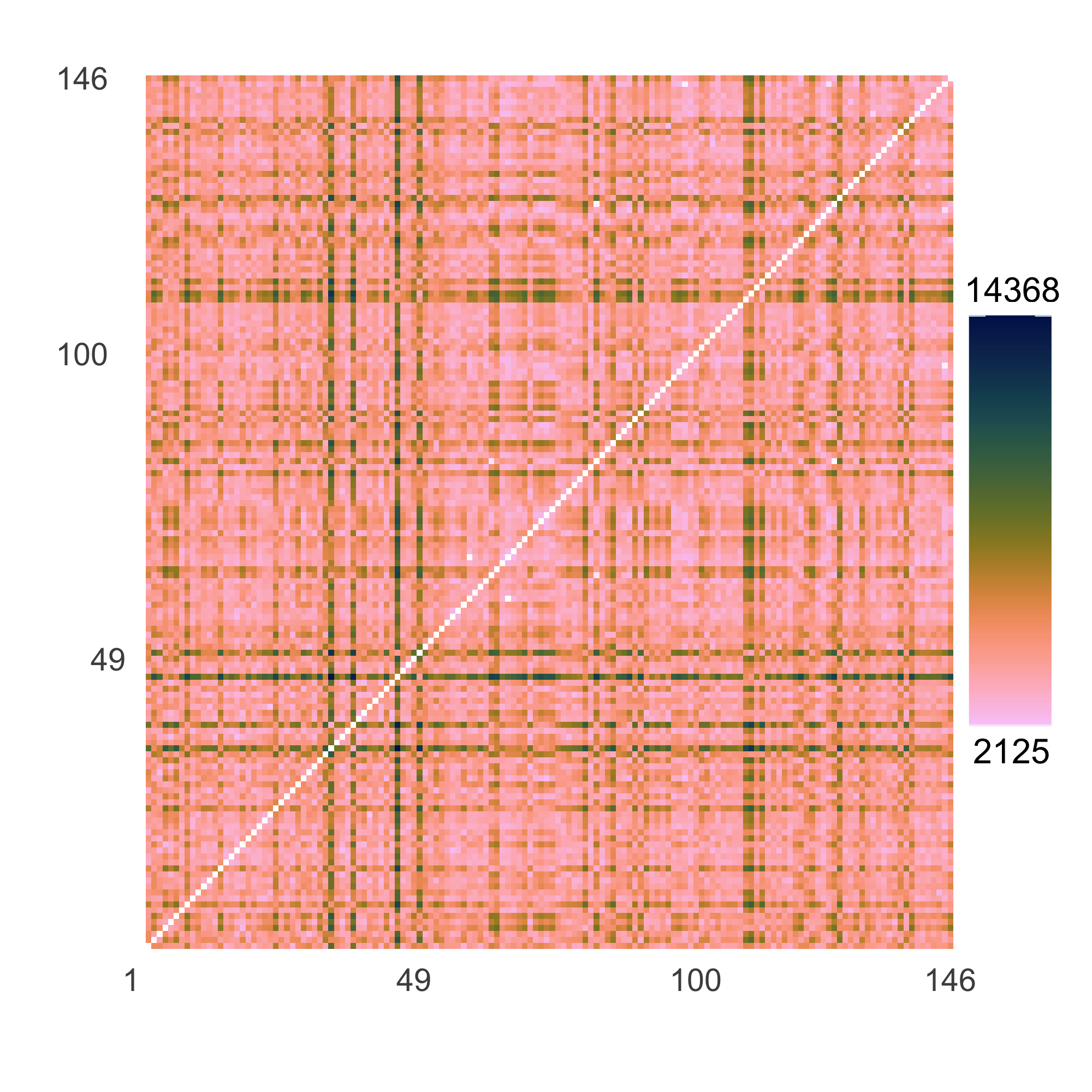}
		\caption{Kendall-Colijn ($\lambda=0$)}
    \end{subfigure}\hfil
	\begin{subfigure}[t]{1.8in}
		\includegraphics[scale=0.06]{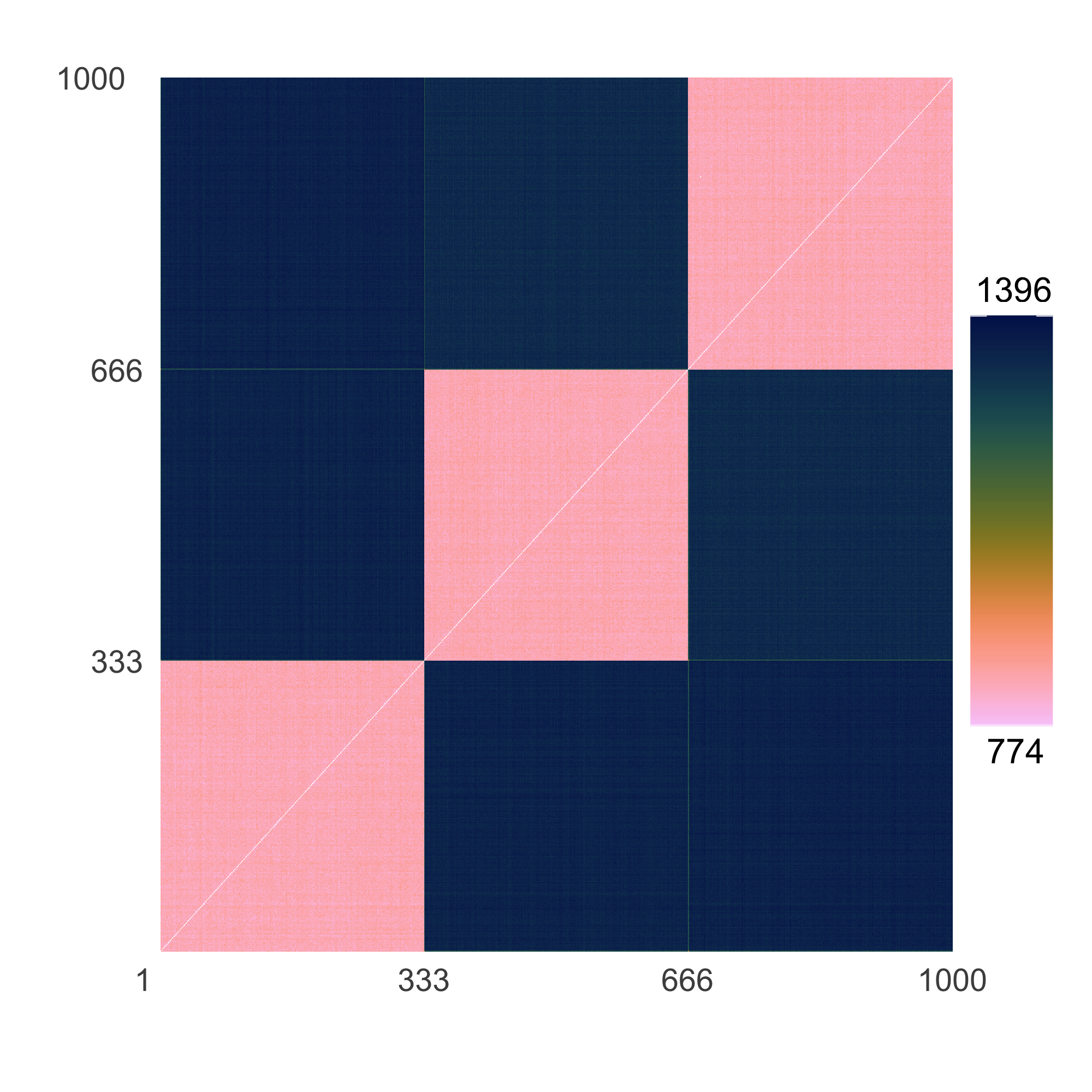}
		\caption{aSPR}
    \end{subfigure}\hfil
	\caption{EBOV heatmaps for all phylogenetic distance metrics. For the Kendall-Colijn distance, the full sample was downsampled to 146 trees equally spaced between tree 1 and tree 1000, as computing pairwise distances for 1000 trees was not feasible.}
	\label{fig:supheatmapsEBOV}
	\end{center}
\end{figure*}

\begin{figure*}[h]
	\begin{center}
		\begin{subfigure}[t]{1.7in}
			\includegraphics[scale=0.3]{figures/NetworkHIV_RF.pdf}
		\caption{Robinson-Foulds}
    \end{subfigure}\hfil
	\begin{subfigure}[t]{1.7in}
		\includegraphics[scale=0.3]{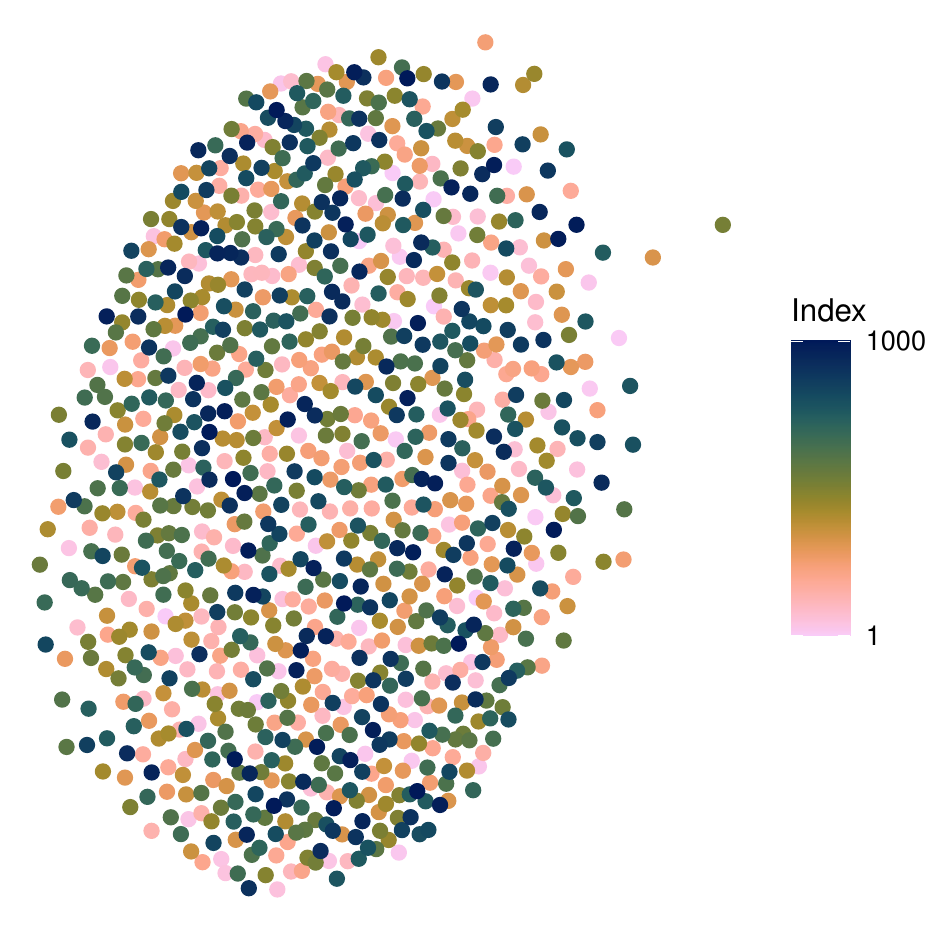}
		\caption{Weighted Robinson-Foulds}
    \end{subfigure}\hfil
	\begin{subfigure}[t]{1.7in}
		\includegraphics[scale=0.3]{figures/NetworkHIV_KF.pdf}
		\caption{Branch score}
    \end{subfigure}\hfil
	\begin{subfigure}[t]{1.7in}
		\includegraphics[scale=0.3]{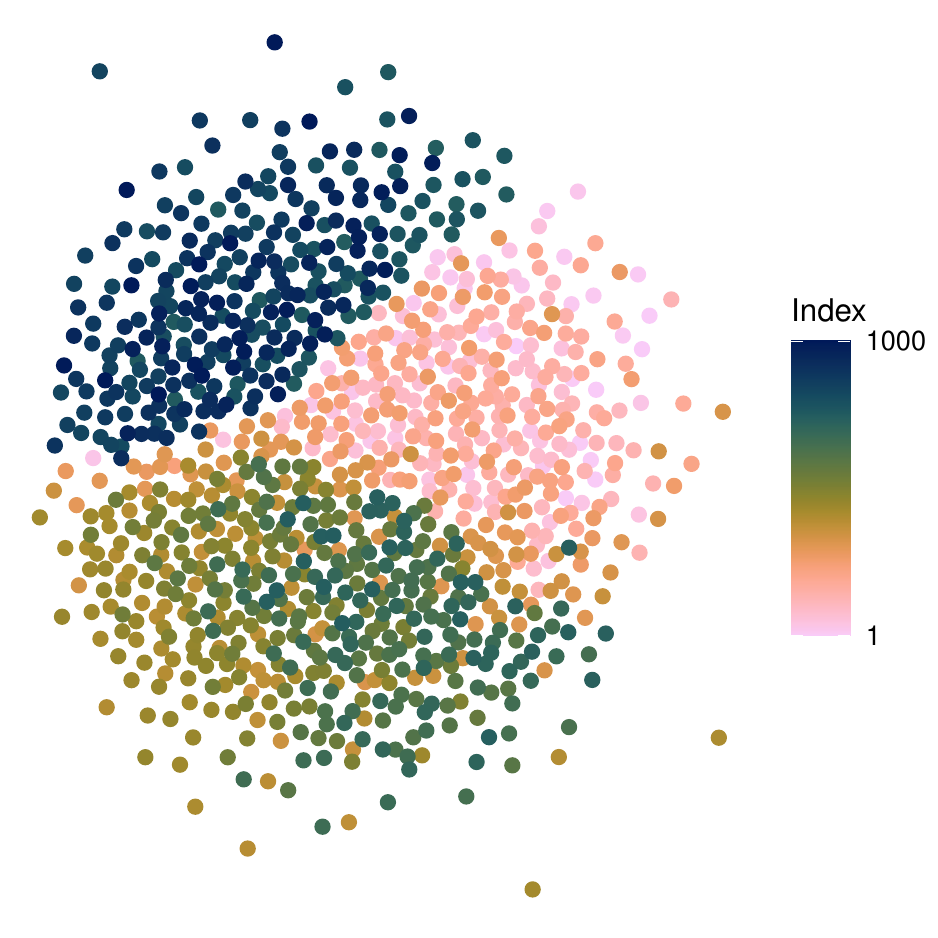}
		\caption{Path difference}
    \end{subfigure}\hfil
	\begin{subfigure}[t]{1.7in}
		\includegraphics[scale=0.3]{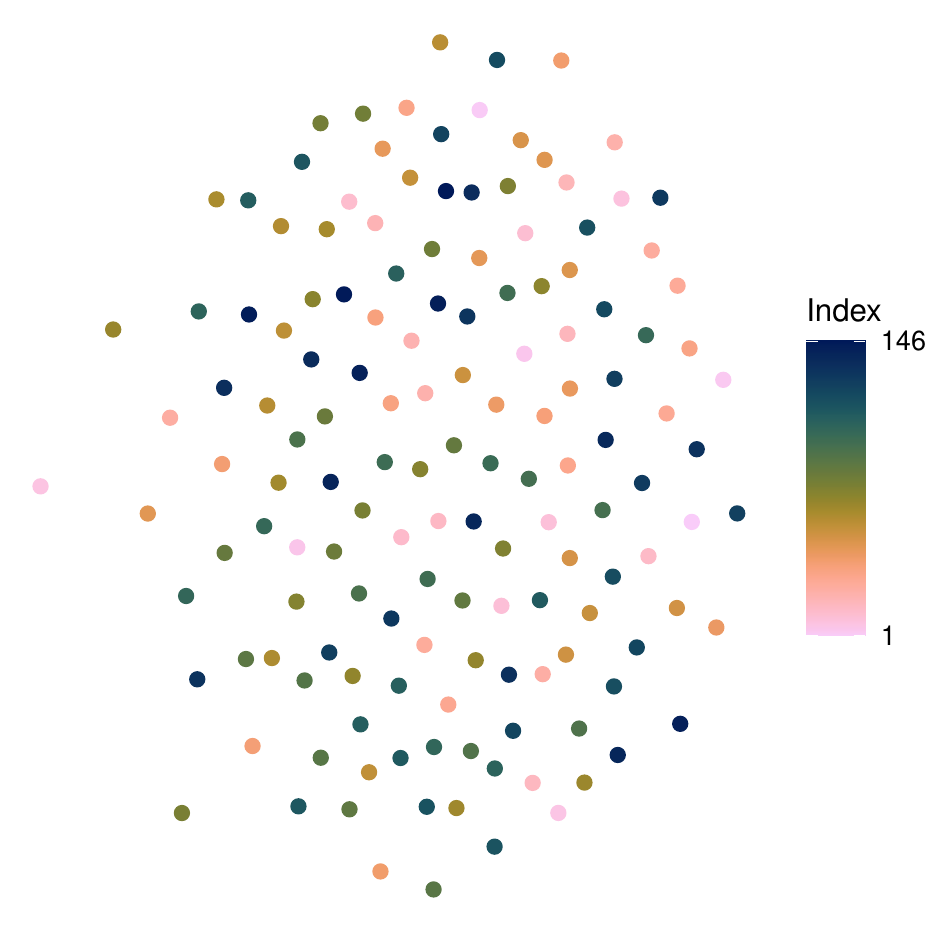}
		\caption{Kendall-Colijn ($\lambda=0$)}
    \end{subfigure}\hfil
	\begin{subfigure}[t]{1.6in}
		\includegraphics[scale=0.3]{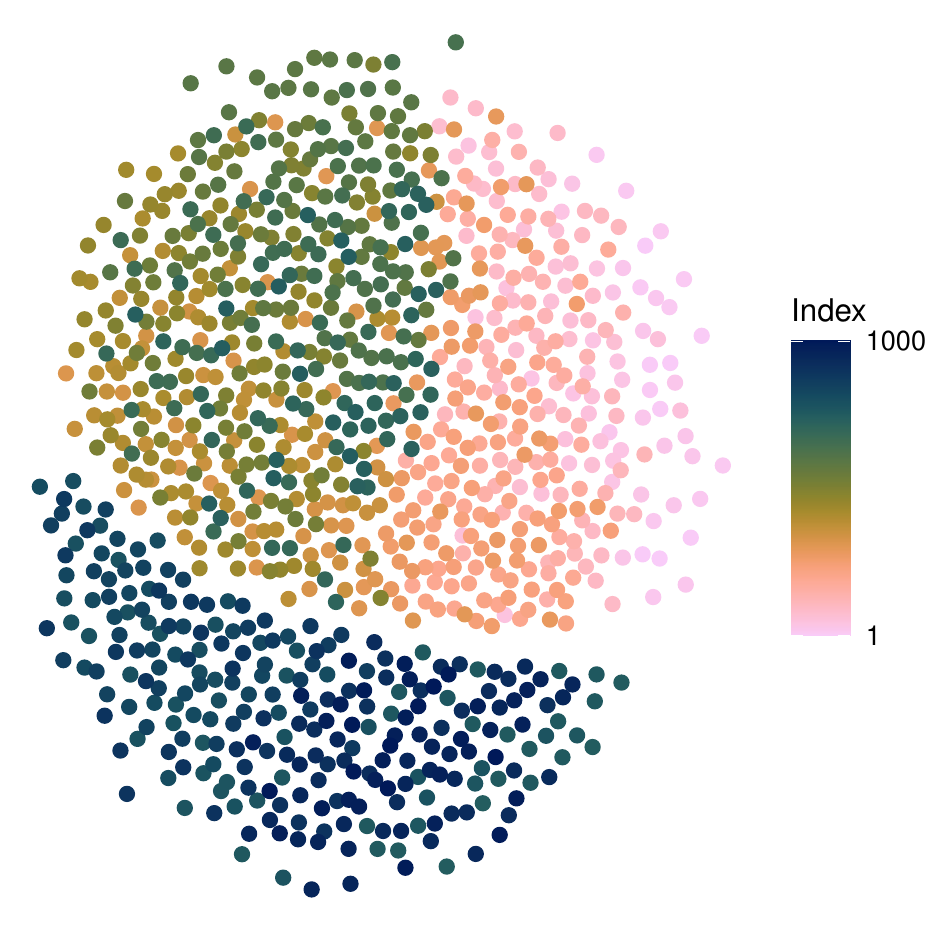}
		\caption{aSPR}
    \end{subfigure}\hfil
	\caption{HIV networks for all phylogenetic distance metrics. For the Kendall-Colijn distance, the full sample was downsampled to 146 trees equally spaced between tree 1 and tree 1000, as computing pairwise distances for 1000 trees was not feasible.}
	\label{fig:supnnetworksHIV}
	\end{center}
\end{figure*}

\begin{figure*}[h]
	\begin{center}
		\begin{subfigure}[t]{1.8in}
			\includegraphics[scale=0.06]{figures/HeatmapHIV_RF.png}
		\caption{Robinson-Foulds}
    \end{subfigure}\hfil
	\begin{subfigure}[t]{1.8in}
		\includegraphics[scale=0.06]{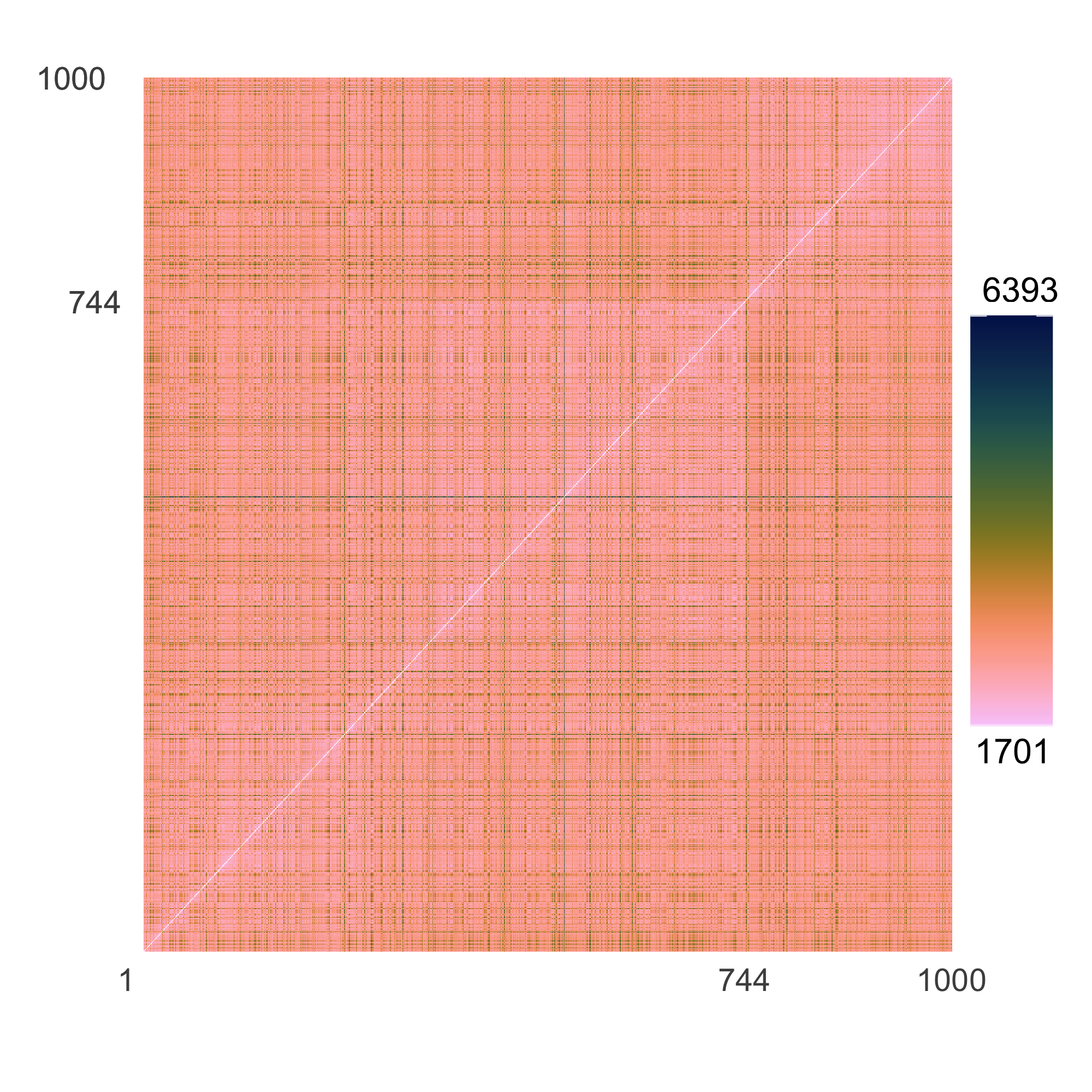}
		\caption{Weighted Robinson-Foulds}
    \end{subfigure}\hfil
	\begin{subfigure}[t]{1.8in}
		\includegraphics[scale=0.06]{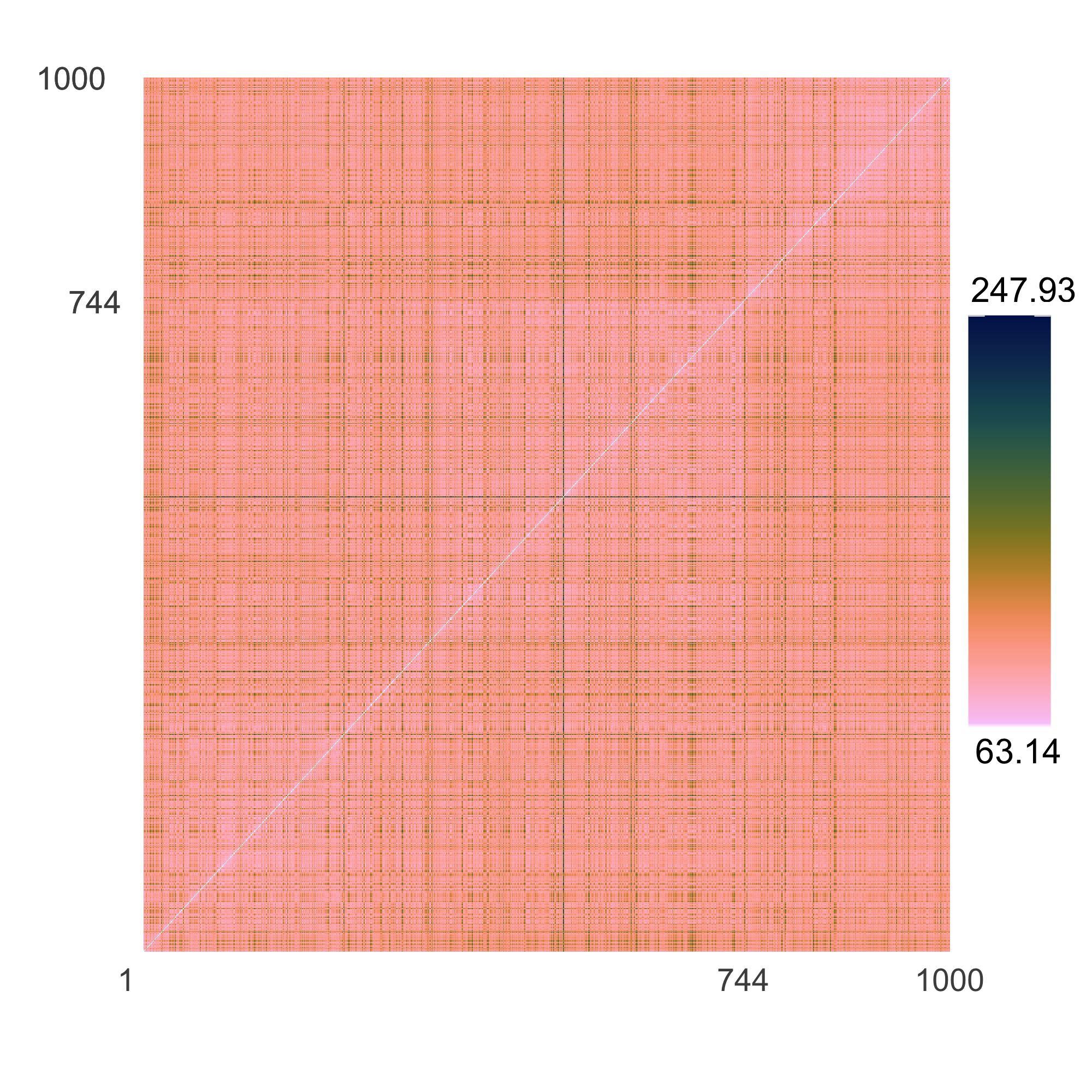}
		\caption{Branch score}
    \end{subfigure}\hfil
	\begin{subfigure}[t]{1.8in}
		\includegraphics[scale=0.06]{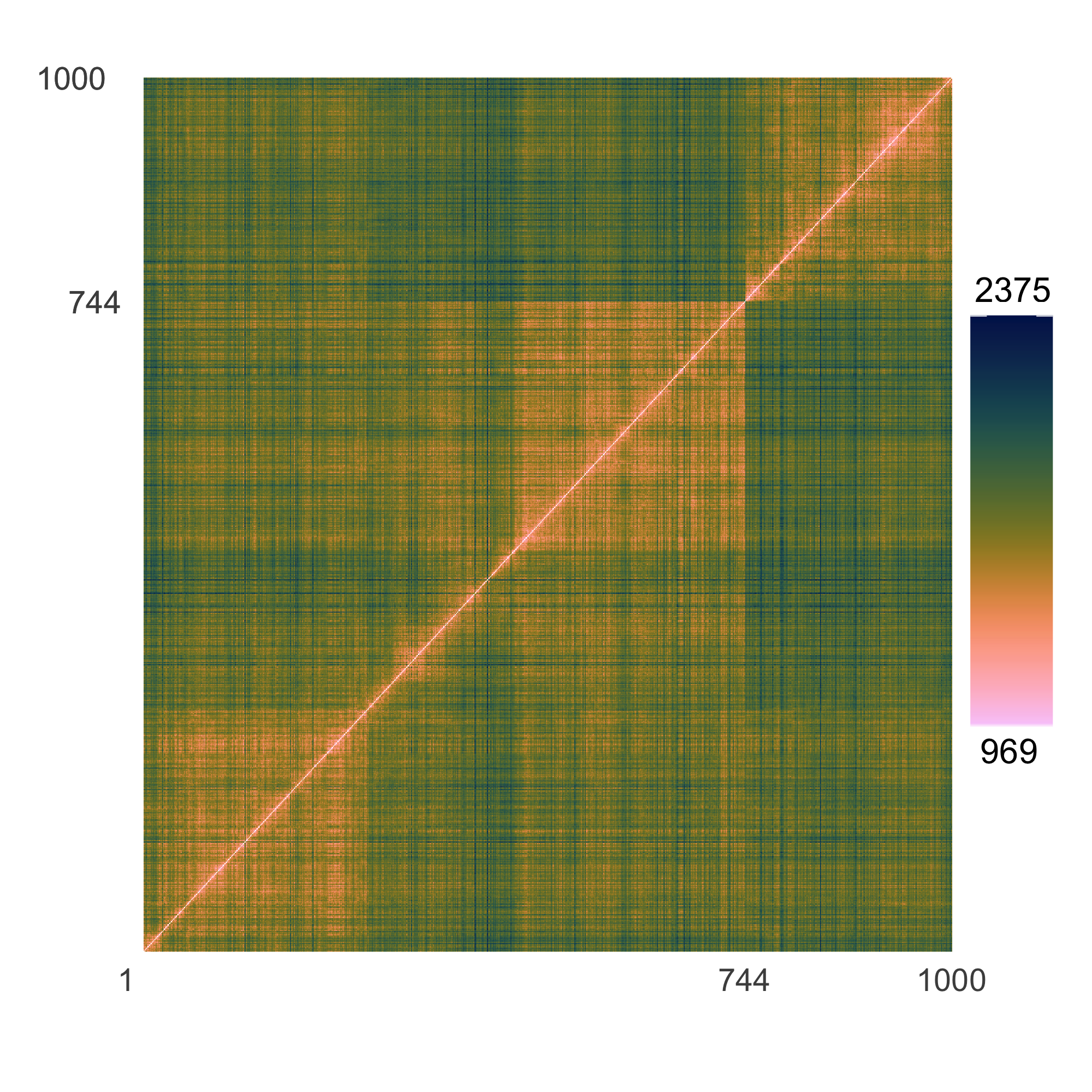}
		\caption{Path difference}
    \end{subfigure}\hfil
	\begin{subfigure}[t]{1.8in}
		\includegraphics[scale=0.06]{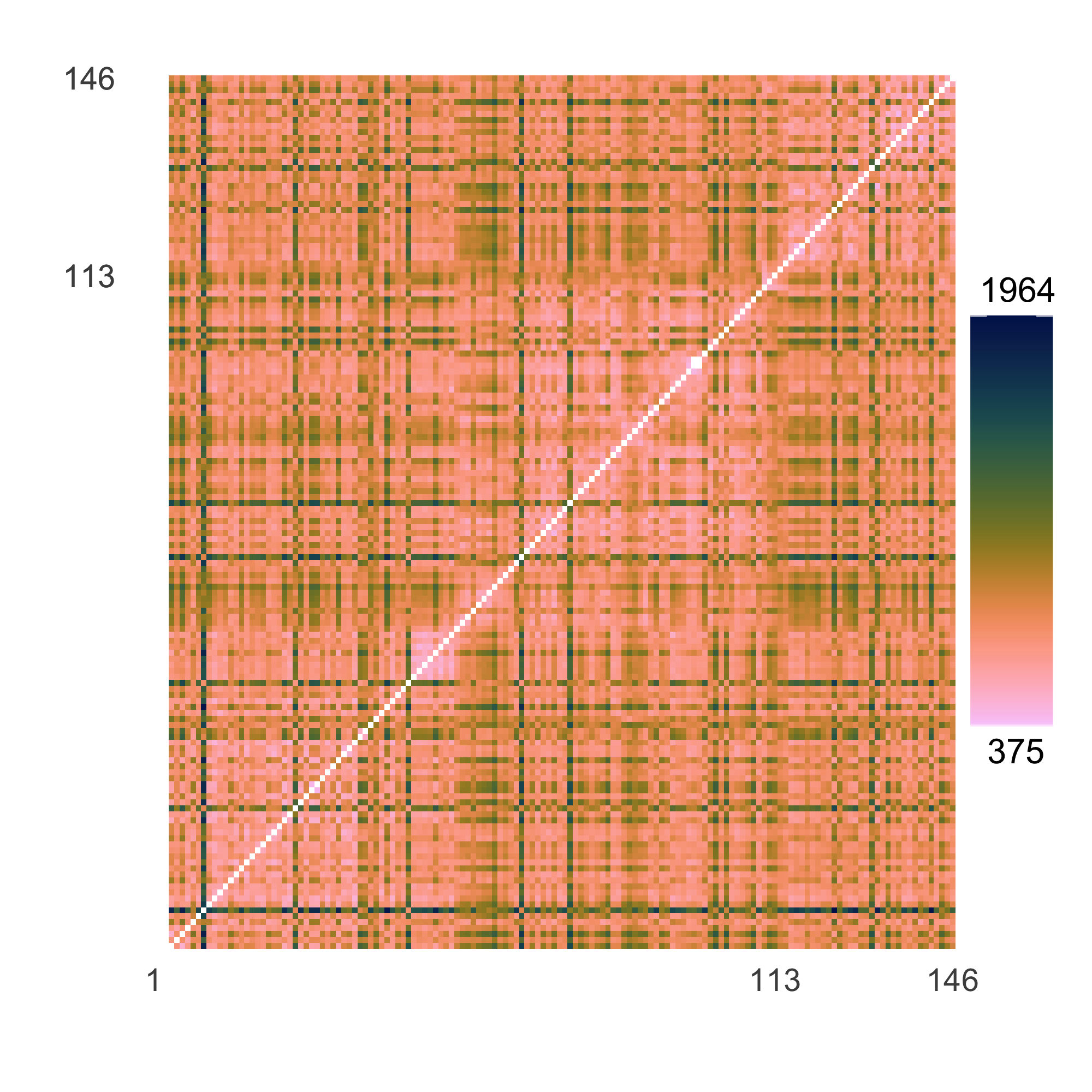}
		\caption{Kendall-Colijn ($\lambda=0$)}
    \end{subfigure}\hfil
	\begin{subfigure}[t]{1.8in}
		\includegraphics[scale=0.06]{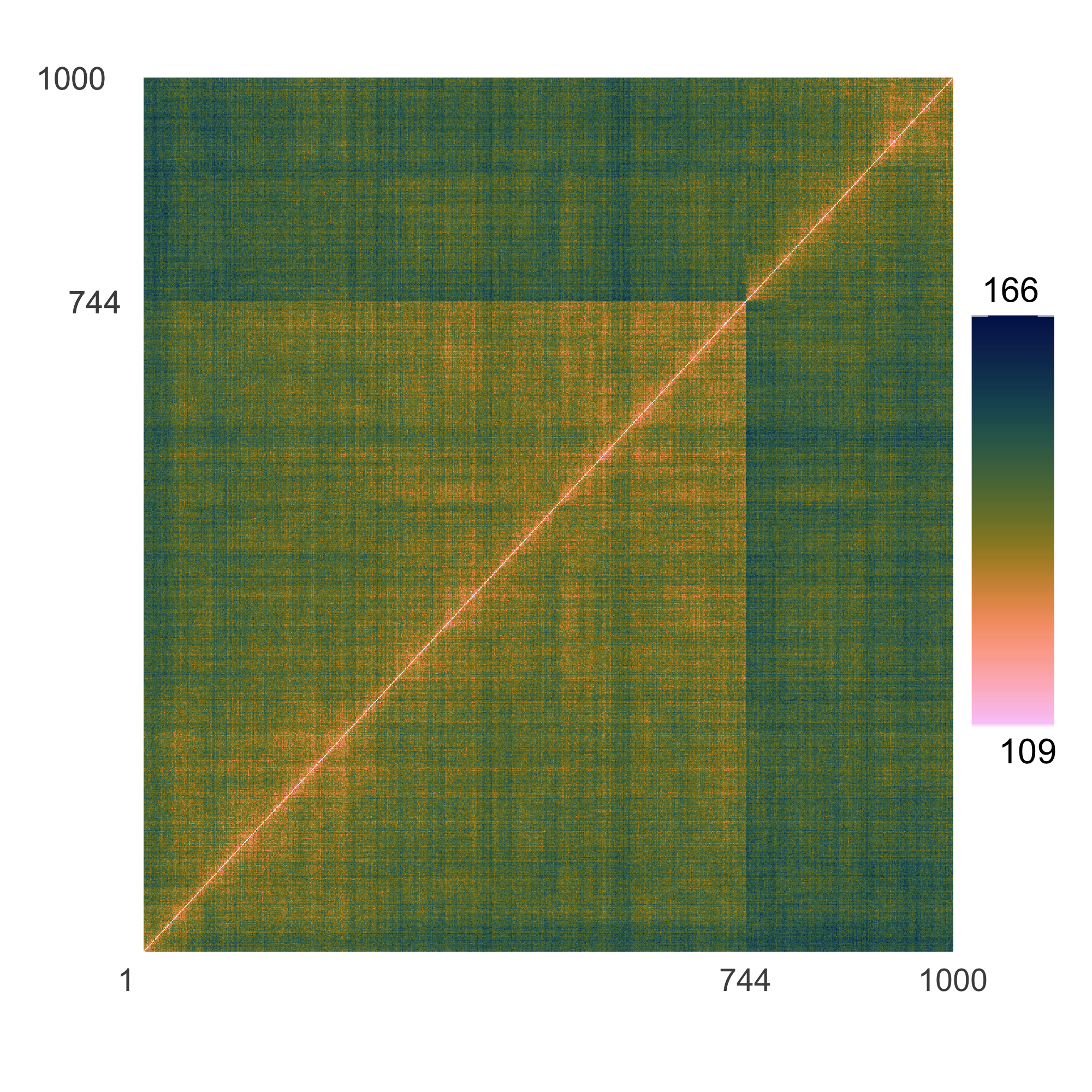}
		\caption{aSPR}
    \end{subfigure}\hfil
	\caption{HIV heatmaps for all phylogenetic distance metrics. For the Kendall-Colijn distance, the full sample was downsampled to 146 trees equally spaced between tree 1 and tree 1000, as computing pairwise distances for 1000 trees was not feasible.}
	\label{fig:supheatmapsHIV}
	\end{center}
\end{figure*}

\begin{figure*}[h]
	\begin{center}
	\begin{subfigure}[b]{3.3in}
		\includegraphics[width=.95\linewidth]{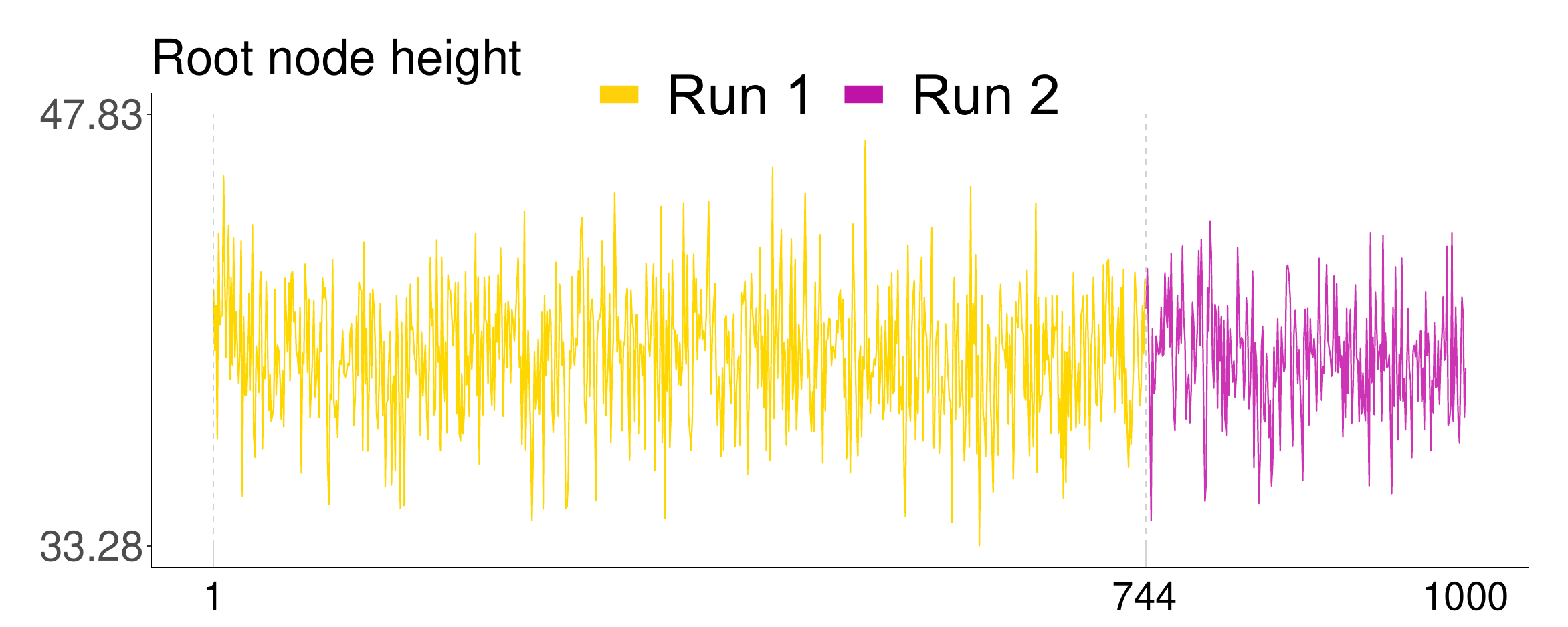}
		\caption{Root node height (HIV trees) --- $ESS=840$}
    \end{subfigure}\hfil
	\begin{subfigure}[b]{3.3in}
		\includegraphics[width=.95\linewidth]{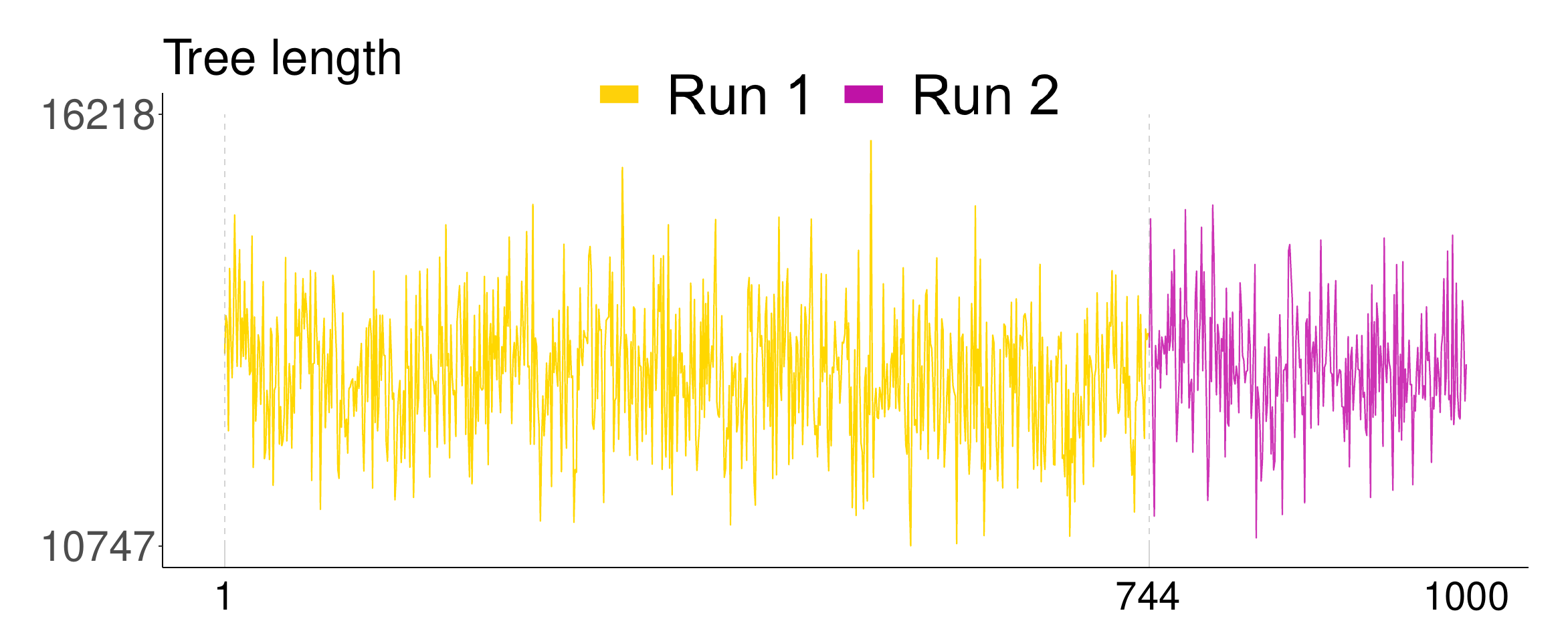}
		\caption{Tree length (HIV trees) --- $ESS=687$}
    \end{subfigure}
	\caption{Trace plots for the root node height and tree length for the HIV samples. Tree length refers to the sum of all tree branches, and is thus a statistic closely related to the topology of the tree. Neither of these statistics show a discrepancy between the two runs. Note that the shape of the traces are extremely similar to each other (although not identical), which could again be related to the star-like shape of HIV phylogenies.}
	\label{fig:suptracerplots}
	\end{center}
\end{figure*}

\begin{figure*}[h]
	\begin{center}
	\includegraphics[scale=0.2]{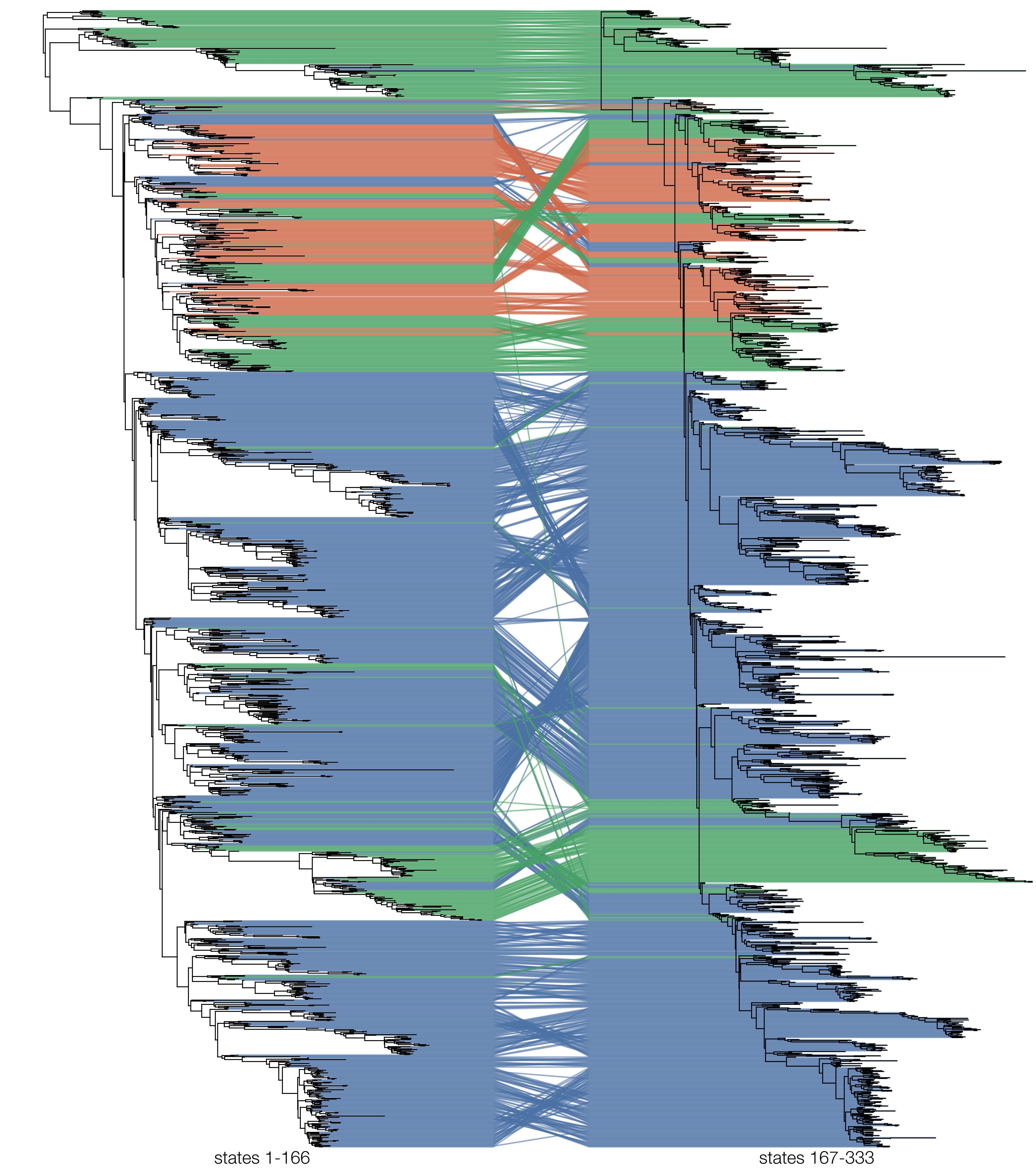}
    \includegraphics[scale=0.4]{figures/legend.pdf}
	\caption{Tanglegram of the MCC trees of the first and second half of the first EBOV run. The tips of the trees are connected to each other by lines, coloured by the country of origin. Disagreement between the subsamples regarding the location of several clades is apparent by the fact that the lines connecting the tips of these clades are not parallel.}
	\label{fig:tangleEBOVsupp}
	\end{center}
\end{figure*}

\begin{figure*}[h]
	\begin{center}
	\includegraphics[scale=0.15]{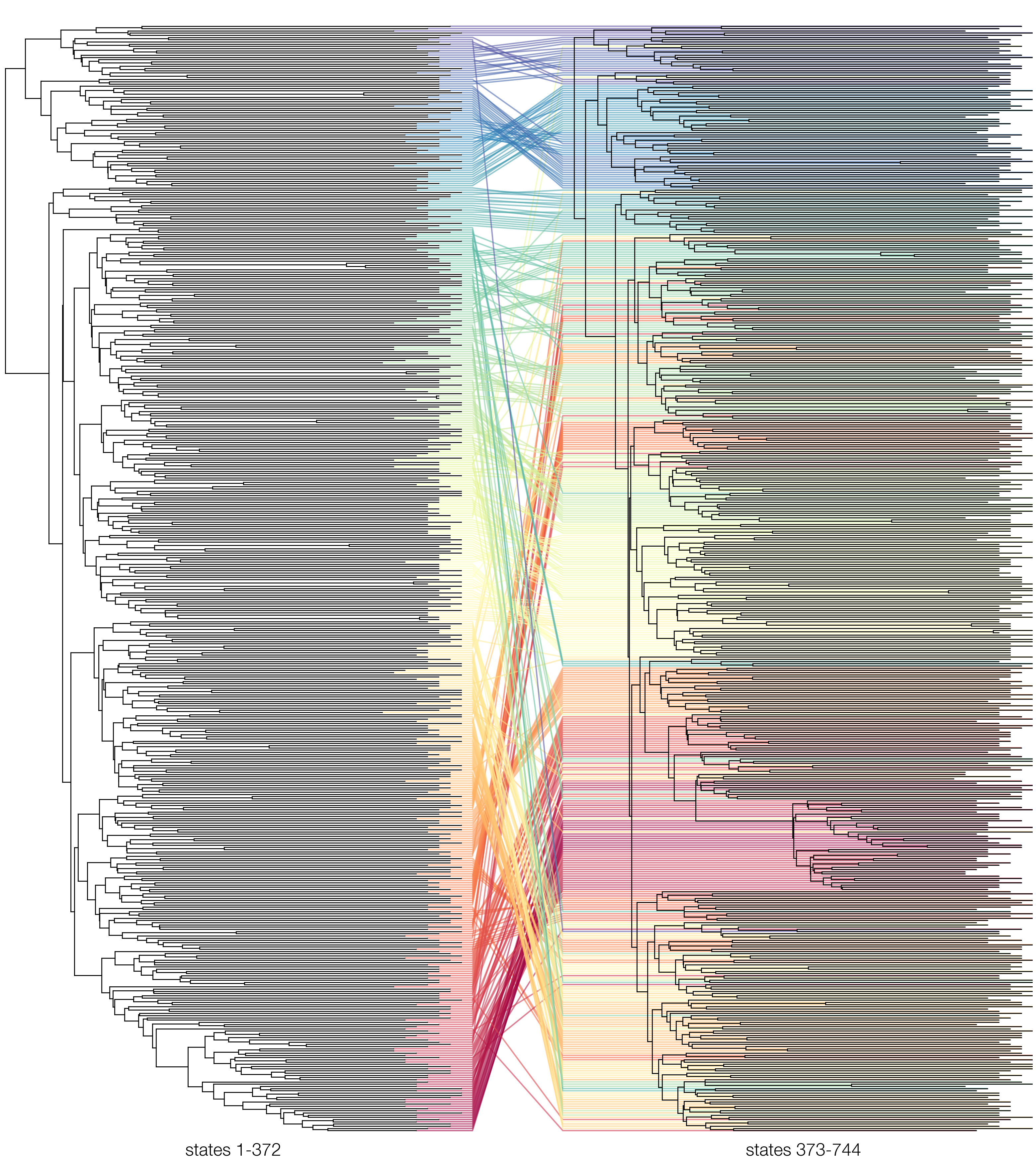}
	\caption{Tanglegram of the MCC trees of the first and second half of the first HIV run. The tips of the trees are connected to each other by lines, coloured by position in the first MCC tree. Disagreement between the subsamples regarding the location of several clades is apparent by the fact that the lines connecting the tips of these clades are not parallel.}
	\label{fig:tangleHIVsupp}
	\end{center}
\end{figure*}

\clearpage

\begin{table}[t]
	\centering
	\caption{Root height and tree length ESS for EBOV subsamples.}
	\begin{tabular}{ccc}
	\toprule
	\textbf{Subsample} & \textbf{Root Height} & \textbf{Tree Length} \\ 
	\midrule
	1-333              & 204                  & 113                  \\ 
	334-666            & 173                  & 140                  \\ 
	667-1000           & 309                  & 143                  \\ 
	1-1000             & 726                  & 20                   \\ 
	\bottomrule
	\end{tabular}
	\label{tab:ebovess}
\end{table}

\begin{table}[t]
	\centering
	\caption{Root height and tree length ESS for HIV subsamples.}
	\begin{tabular}{ccc}
	\toprule
	\textbf{Subsample} & \textbf{Root Height} & \textbf{Tree Length} \\ 
	\midrule
	1-744              & 609                  & 366                  \\ 
	745-1000           & 221                  & 220                  \\ 
	1-1000             & 840                  & 687                  \\ 
	\bottomrule
	\end{tabular}
	\label{tab:hivess}
\end{table}

\end{document}